\documentclass[letterpaper,10pt, oneside]{article}
\usepackage{stmaryrd}
\usepackage{latexsym, amsmath, amssymb, amsfonts, amscd}
\usepackage{bbm}
\usepackage{amsxtra}
\usepackage[T1]{fontenc}
\usepackage{lmodern}
\usepackage{nicefrac,mathtools,enumitem}
\usepackage[mathscr]{eucal}
\usepackage{pgf,calc,pgffor,xkeyval}
\usepackage[all]{xy}
\usepackage{tikz}
\newtheorem{thm}{Theorem}[section]
\newtheorem{lem}[thm]{Lemma}\newtheorem{lemma}[thm]{Lemma}
\newtheorem{cor}[thm]{Corollary}
\newtheorem{pro}[thm]{Proposition}
\newtheorem{pdef}[thm]{Proposition-Definition}
\newtheorem{ex}[thm]{Example}
\newtheorem{ep}[thm]{Example}
\newtheorem{rmk}[thm]{Remark}\newtheorem{remark}[thm]{Remark}

\newtheorem{defi}[thm]{Definition}

\newcommand{\lon }{\,\rightarrow\,}
\newcommand{\be }{\begin{equation}}
\newcommand{\ee }{\end{equation}}

\newcommand{\pf}{\noindent{\bf Proof.}\ }

\newcommand{\R}{\mathbb R}\newcommand{\Z}{\mathbb Z}

\newcommand{\huaA}{\mathcal{A}}

\newcommand{\huaE}{\mathcal{E}}
\newcommand{\huaF}{\mathcal{F}}
\newcommand{\huaG}{\mathcal{G}}

\newcommand{\huaV}{\mathcal{V}}

\newcommand{\huaT}{\mathcal{T}}

\newcommand{\huaM}{\mathcal{M}}

\newcommand{\CWM}{C^{\infty}(M)}

\newcommand{\frka}{\mathfrak a}

\newcommand{\frkg}{\mathfrak g}\newcommand{\g}{\mathfrak g}
\newcommand{\frkh}{\mathfrak h}

\newcommand{\frkX}{\mathfrak X}

\def\qed{\hfill ~\vrule height6pt width6pt depth0pt}

\newcommand{\half}{\frac{1}{2}}

\newcommand{\pair}[1]{\left\langle #1\right\rangle}

\newcommand{\Courant}[1]{\left\llbracket  #1\right\rrbracket }

\newcommand{\p}{\mathbbm{p}}
\newcommand{\inc}{\mathbbm{i}}

\newcommand{\Hom}{\mathrm{Hom}}
\newcommand{\Der}{\mathrm{Der}}

\newcommand{\Sym}{\mathrm{Sym}}

\newcommand{\End}{\mathrm{End}}
\newcommand{\ad}{\mathrm{ad}}

\newcommand{\inv}{\mathrm{inv}}

\newcommand{\ve}{\mathrm{v}}
\newcommand{\h}{\mathrm{h}}
\newcommand{\sgn}{\mathrm{sgn}}
\newcommand{\Ksgn}{\mathrm{Ksgn}}

\newcommand{\hl}{\hat{l}}
\newcommand{\hrho}{\hat{\rho}}
\newcommand{\hA}{\widehat{A}}
\newcommand{\hg}{\hat{\frkg}}

\newcommand{\tF}{\widetilde{F}}
\newcommand{\td}{\widetilde{\partial}}
\newcommand {\comment}[1]{{\marginpar{*}\scriptsize\textbf{Comments:}
    #1}}
\newcommand {\emptycomment}[1]{}

\newcommand{\GpdBibd}{{\bf GpdBibd}}
\newcommand{\Gpd}{{\bf Gpd}}

\DeclareMathOperator{\im}{im}

\begin{document}
\title{
{Higher Extensions of Lie Algebroids\thanks
 {This research is supported by NSF of China (11471139), NSF of Jilin Province (20140520054JH) and the German Research Foundation
(Deutsche Forschungsgemeinschaft (DFG)) through the Institutional
Strategy of the University of G\"ottingen.
 } } }
\author{Yunhe Sheng  \\
Department of Mathematics, Jilin University,\\
 Changchun 130012,  China
\\\vspace{3mm}
email: shengyh@jlu.edu.cn\\
Chenchang Zhu\\
Courant Research Center ``Higher Order Structures'',\\
Georg-August-University
G$\rm\ddot{o}$ttingen, \\Bunsenstrasse 3-5, 37073, G$\rm\ddot{o}$ttingen, Germany\\
email:zhu@uni-math.gwdg.de }

\maketitle 


\begin{abstract}
We study the extension of a Lie algebroid by a representation up to
homotopy, including   semidirect products of a Lie
algebroid with such representations. The extension results in  a higher
Lie algebroid. We give exact Courant algebroids
and string Lie 2-algebras as examples of such extensions. We then
apply this to obtain a Lie 2-groupoid integration of an exact Courant algebroid.
\end{abstract}
\tableofcontents

\section{Introduction}
In this paper we study the abelian extension of a Lie algebroid $A$ by a sort of
higher representation. This concept of higher
representation of a Lie algebroid is an extension of
Lada-Markl's $L_\infty$-modules \cite{LadaMartin} to the context of Lie algebroids. It has been developed recently by Abad and
Crainic under the name of
representations up to homotopy \cite{abad-crainic:rep-homotopy} and by
Gracia-Saz and   Mehta under the name of superconnection
\cite{mehtaa}. The latter one and a second article \cite{mehtab} by the same authors lead further to Mehta and Tang's integration result \cite{MT} of a Courant algebroid that we will mention later.

Recently,  representations up to homotopy have been widely
studied from various aspects. For example, they are studied from the
aspect of equivariant cohomology in \cite{abad-crainic:rep-gpd},
deformation theory in \cite{abad-schaetz:deform},
integration theory in \cite{abad-schaetz:integration-rep-homotopy}, VB-algebroids and VB-groupoids in \cite{Jotz, mehtab},
double Lie algebroids in \cite{GJMM}, Lie algebroid
modules  and  modular classes in \cite{Meh, Meh2},
 $L_\infty$-algebra actions in \cite{MZ}, and degree 2 Poisson
 N-manifolds and Poisson Lie 2-algebroids  in \cite{Jot}.

 To
talk about higher
extensions of a Lie algebroid, we first bring the concept of a split Lie $n$-algebroid onto
the surface (Definition
\ref{def:n-algd}). It is then further specially studied in the work of \cite{BP,Jot}. Some concepts, such as Courant algebroids, can be
described using the language of
differential geometry as in \cite{lwx}. However, the
description of them via  NQ manifolds better reflects their nature, though that
perspective often involves calculations in local
coordinates. Split Lie $n$-algebroids which we introduce
represent a middle way, which  allows us
to study NQ manifolds within the framework of differential geometry.
An NQ-manifold is a non-negatively
graded manifold $\huaM$ together with a degree $1$ vector field $Q$
satisfying $[Q,Q]=0$.  A degree 1 NQ-manifold is a  Lie
algebroid. In fact, a degree 1 non-negatively graded
manifold can be modeled by a vector bundle with shifted degree
$A[1]\to M$. The function ring of $A[1]$ is the graded algebra
\begin{equation}\label{eq:func-ring-algd}
 C(A[1]) = C^\infty(M) \oplus \Gamma( A^*) \oplus \Gamma(\wedge^2 A^*) \oplus \cdots \end{equation}
A degree 1 vector field $Q$ is a degree 1 derivation of this
algebra. Equivalently, this means that we have a vector bundle
morphism (called the  anchor later on) $\rho_A: A \to TM$ and a
Lie bracket $[\cdot,\cdot]_A$ on $\Gamma(A)$ such that $Q = d_A$,
where $d_A: \Gamma(\wedge^{n} A^*) \to \Gamma(\wedge^{n+1} A^*)$ is
defined as the generalized de Rham differential 
\begin{equation} \label{eq:da}
 d_A (\xi) (X_0,\dots,X_n) =  \sum_{i<j} (-1)^{i+j}\xi([X_i, X_j]_A, \dots,\hat{X}_i,\dots,\hat{X}_j,\dots) + \sum_{i=0}^n (-1)^{i} \rho_A(X_i)\xi(\dots,\hat{X}_i,\dots).
 \end{equation}
The equation $[Q, Q]=0$ is then equivalent to the condition required for $(A, \rho_A, [\cdot, \cdot]_A)$ to be a Lie algebroid,
 that is, $[[X,Y]_A,Z]_A+c.p.=0$ and $ [X, fY]_A= f[X, Y]_A+ \rho_A (X)(f)Y$  for all $X, Y, Z\in \Gamma(A)$ and $f\in
C^\infty(M)$.
There is also another method to recover the Lie bracket on $A$:
$\Gamma(A)$ can be viewed as the space of degree $-1$ vector fields
on $A[1]$.
  Then a degree 1 homological vector field $Q$ on $A[1]$
gives us a derived bracket $[X, Y]_A:= [[Q, X], Y]$, which is
exactly the Lie algebroid bracket on $A$ corresponding to $Q$ given
above.

Now we apply  the same procedure for higher NQ manifolds and we encounter a different story. First of all, to model a degree
$n$ non-negatively graded manifold by a graded vector bundle requires
an unnatural choice of connection (even though it is always
possible). It is comparable to the fact that to single out a
composition of 1-cells of  a Lie $n$-groupoid $X_\bullet$ modeled
using a Kan simplicial manifold requires an unnatural choice of a
section from the horn space $X_1 \times_{X_0} X_1$ to $X_2$ (in this
case it is  not always possible\footnote{However it is always
  possible locally which explains the existence of the choice in the
  infinitesimal case. (Private conversation with Dimitry
  Roytenberg.)}). However, there are still some circumstances in
  which
a graded vector bundle arises naturally (namely a preferred
connection is chosen somehow); for example, extensions and semidirect products of a Lie
algebroid with its higher representations (see the next paragraph). For this reason we will assume that our
degree $n$ N-manifold comes from a graded vector bundle $\huaA = A_0
\oplus A_{-1} \oplus \dots \oplus A_{-n+1}$. Then, similarly to
\eqref{eq:func-ring-algd}, the function ring is the graded
commutative algebra,
\begin{equation} \label{eq:func-ring}
\begin{split}
C(\huaA[1])&:= C^\infty(M) \oplus \Gamma(\Sym(\huaA[1])^*) \\
                    &=C^\infty(M)\oplus\Big[\Gamma(A_0^*)\Big]\oplus
 \Big[\Gamma(\wedge^2A_{0}^*)\oplus\Gamma(A_{-1}^*)\Big]\oplus\Big[\Gamma(\wedge^3A_0^*)\oplus\Gamma(A_0^*)\otimes\Gamma(A_{-1}^*)\oplus\Gamma(A_{-2}^*)\Big]\oplus\cdots.
\end{split}
\end{equation} where $A_{-i}^*$ has degree $i+1$ and $C^\infty(M)$ lies
in degree 0. Then a homological degree 1 vector field $Q$ gives us
an anchor $\rho$ and various brackets $l_i$ for $i=1, \dots, n+1$
(see Definition \ref{def:n-algd}). We call such an object a {\bf split Lie
$n$-algebroid}. For example, the NQ
manifold $T^*[2]T[1]M$ corresponding to an
exact Courant algebroid has an anchor
$\rho$ and various brackets $l_i$  given by
\eqref{eq:ext-a-structure}. The
Courant bracket \eqref{eqn:bracket} arises as the derived bracket, which is, unlike
the degree 1 case,
{\bf different} from the 2-bracket $l_2$, no matter how we choose the
splitting. It seems that, recently, the language of split Lie $n$-algebroids has slowly become a useful tool for differential geometers to study problems related to NQ manifolds \cite{BP, Jot}.

Once we clarify what higher Lie algebroids are, the concept of a
semidirect product  of $A$ with a higher representation $\huaE$ is
naturally defined (see Proposition-Definition \ref{lem:ruthLie2a}). Following this, we
 study  extensions. We obtain a   classification result of the 2-term abelian extensions of a Lie
algebroid $A$ by $H^2(A, \huaE)$, where $\huaE$ is a 2-term
representation up to homotopy of $A$ (Theorem \ref{thm:ext-a}). The
result is   totally analogous to the classical case.
\emptycomment{: cocycles
representing the same
class in $H^2(A, \huaE)$ give  isomorphic 2-term
abelian extensions; however we can show that the converse is true only
in some specific situations involving the property of the boundary map
in $\huaE$. The obstruction in the proof exists even for central
extensions of 2-term
$L_\infty$-algebras.  This is because the cohomology
group $H^\bullet (\huaA, \huaE)$ we use is taking simply as the cohomology of
the complex $D: C(\huaA)\otimes \Gamma(\huaE)\to C(\huaA)\otimes
\Gamma(\huaE) $. } In the case of extensions of $L_\infty$-algebras, Lazarev gave a full version of one-to-one correspondence in \cite{lazarev:extension}. See Remark \ref{rmk:cohomology} for details.
\emptycomment{
This probably relates to the fact that
$L_\infty$-algebras and (non-strict) $L_\infty$-morphisms lack
colimits and limits, and (co)limits sometime have to be taken in a
homotopic fashion.  So the classical proof for Lie algebras breaks
down\footnote{Private talk to
  Bruno Vallette,  Ezra Getzler and Urs Schreiber.} when applied to
$L_\infty$-algebras.
}
   It follows  that
the semidirect product (the trivial extension) is in bijective  correspondence with the
$0$-class in $H^2(A, \huaE)$, which justifies the definition of
a semidirect product.

Examples of such extensions include  Courant
algebroids, their higher analogue,  and string Lie 2-algebras. Courant algebroids (see Section \ref{sec:courant}) were first
introduced in \cite{lwx} to study doubles of Lie bialgebroids. An
equivalent definition via graded manifolds was given by Roytenberg
in \cite{Roytenbergphdthesis}. Then Roytenberg and Weinstein discovered the
relationship between Courant algebroids and $L_\infty$-algebras \cite{rw},
which was an indication of the higher structures behind Courant
algebroids.  Here we briefly recall that an exact Courant algebroid
$TM\oplus T^*M$ with the \v{S}evera class $[H]$, where $H\in
\Omega^3(M)$, has the antisymmetric bracket
\begin{equation}\label{eqn:bracket}
\Courant{X+\xi,Y+\eta}\triangleq[X,Y]+L_X\eta-L_Y\xi+\frac{1}{2}d(\xi(Y)-\eta(X))+i_{X\wedge
Y}H.
\end{equation} When $H=0$, this defines the standard Courant
algebroid. We show that the Lie
2-algebroid underlying the standard Courant algebroid  is the semidirect product $TM\ltimes (T^*M\xrightarrow{id}
T^*M)$. Its higher analogue $T^*[r]T[1]M$ as a Lie $r$-algebroid is
the semidirect product $TM \ltimes (T^*M \xrightarrow{id} T^*M \to  0
\dots\to 0)$. Moreover, the $L_\infty$-algebras corresponding to such
Courant algebroids constructed in \cite{rw, zambon:l-infty,
  getzler:higher-derived} are all  semidirect products of a
Lie algebra with a representation up to homotopy.  Now an exact Courant algebroid with the \v{S}evera class $[H]$
is an extension with the extension class determined by $H$,
\[0\longrightarrow \big( T^*M \xrightarrow{id} T^*M \big) \longrightarrow \text{Courant algebroid} \longrightarrow TM\longrightarrow 0. \]  However it is easy
to show that
$H^{k\ge 1}(TM, T^*M\xrightarrow{id}
T^*M)=0$ (Lemma \ref{lemma:trivial-coho}), so the Lie 2-algebroid underlying an exact Courant
algebroid is isomorphic   to the one underlying the standard Courant
algebroid regardless the \v{S}evera class.

On the other hand, we do
have a nontrivial extension in the example of a string Lie
2-algebra. A string Lie 2-algebra \cite{baez:string,
  henriques} is the infinitesimal counterpart of
Witten's string Lie 2-group. We realize it as a central
extension of a semisimple Lie algebra $\g$ of compact type by the
abelian Lie 2-algebra $\R\to 0$ (see Example \ref{ep:string}).  This
is not surprising since string Lie 2-groups are central extensions by
$BS^1$.

The framework of extensions helps us to do integration: given an
extension of Lie algebras $\frka \to \hg\to \g$, if we know that
group $G$ integrates $\g$ and we know the integrated extension class
in $H^2(G, \frka)$, then we can write down the group integrating
$\hg$ explicitly. Using this principle, we find group-like objects
to integrate Courant algebroids. The Lie 2-groupoid we find to
integrate a standard Courant algebroid is
\begin{equation}\label{eq:integration-gpd}
\xymatrix{\Pi_1(M)^{\times 2}\times_{M^{\times 3}} (T^*M)^{\times 3}
  \ar@<.7ex>[r] \ar@<-.7ex>[r] \ar[r] & \Pi_1(M)\times_{M} T^*M
  \ar@<.5ex>[r] \ar@<-.5ex>[r] & M,   }
\end{equation}

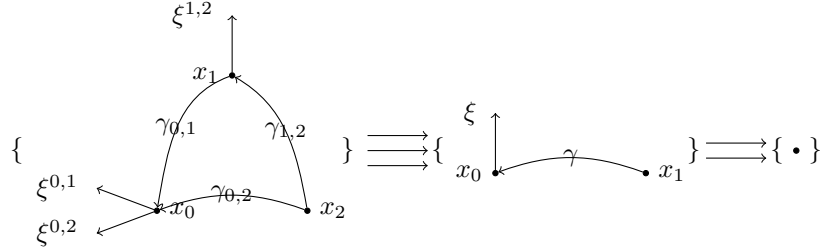
\begin{center}
\begin{tikzpicture}
\draw (-2.2,0.8) node [label=right:\{] {};

\draw (0,0) node [draw,circle,fill=black,minimum size=2pt, inner
sep=0pt] (0) [label=right:$x_0$] {}
 ++(0:2cm) node [draw,circle,fill=black,minimum size=2pt, inner
sep=0pt] (2) [label=right:$x_2$] {}; \draw (1,1.8) node
[draw,circle,fill=black,minimum size=2pt, inner sep=0pt] (1)
[label=left:$x_1$] {};

\draw[<-](0) to [out=80,in=200] node{$\gamma_{0,1}$}
(1);\draw[<-](0) to [out=20,in=160] node{$\gamma_{0,2}$} (2);
\draw[<-](1) to [out=-30,in=100] node{$\gamma_{1,2}$} (2);

 \draw[->]
(0) to (-0.8,0.3) node
 [label=left:$\xi^{0,1}$] {};\draw[->]
(0) to (-0.8,-0.3) node
 [label=left:$\xi^{0,2}$] {};
 \draw[->]
(1) to (1,2.6) node
 [label=left:$\xi^{1,2}$] {};
 \draw (2.2,0.8) node [label=right:\}] {};

 \draw[->] (2.8,1) to (3.6,1);
 \draw[->] (2.8,0.8) to (3.6,0.8);
 \draw[->] (2.8,0.6) to (3.6,0.6);

\draw (3.4,0.8) node [label=right:\{] {};

 \draw(4.5,0.5) node
[draw,circle,fill=black,minimum size=2pt, inner sep=0pt] (0)
[label=left:$x_0$] {}
 ++(0:2cm) node [draw,circle,fill=black,minimum size=2pt, inner
sep=0pt] (1) [label=right:$x_1$] {}; \draw[<-](0) to [out=20,in=160]
node{$\gamma$} (1); \draw[->] (0) to (4.5,1.3) node
 [label=left:$\xi$] {};

 \draw(6.8,0.8) node [label=right:\}] {};

\draw[->] (7.3,0.9) to (8.1,0.9);
 \draw[->] (7.3,0.7) to (8.1,0.7);

 \draw(8.5,0.8) node [draw,circle,fill=black,minimum size=2pt, inner
sep=0pt] (0) [label=right:\}][label=left:\{]{};
\end{tikzpicture}
\end{center}
where $\Pi_1(M)=\tilde{M}\times \tilde{M}/\pi_1(M)$ is the
fundamental groupoid of $M$ with $\tilde{M}$ the simply connected
cover  of $M$. This 2-groupoid is the same (locally) as the one given in recent parallel
works \cite{MT, LS} on integrating Courant algebroids. The work of  \cite{Lie rackoids}
gives a different approach to integrate the Dorfman bracket in a Courant algebroid, which is also very interesting.

We must mention that in this paper we do not  perform  rigorous
differentiation partially because this is the merit
of \cite{LS} (only in Remark \ref{rk:diff} we sketch a possible
differentiation method), and we do not integrate the symplectic form
on a Courant algebroid. Integrating the symplectic form systematically
from our point of view
involves integrating morphisms of higher NQ manifolds, which we do not
deal with here. In their second paper on this topic \cite{MTb}, Mehta
and Tang apply the general Sullivan-\v{S}evera-Getzler-Henriques
integration procedure and obtain a universal integration object which
is an infinite dimensional Lie 2-groupoid (called {\rm LWX} 2-groupoid). As for the symplectic form,
they perform a Cattaneo-Felder style of integration and arrived at a
symplectic form on {\rm LWX} 2-groupoid.
 However, they claim that the relation between the
infinite dimensional model and the above finite dimensional one is not
yet clear.

Last but not least, we mention that general (not
necessarily abelian) extensions,   by usual representations (not
higher representations) of Lie algebroids, and their integrations are studied in
\cite{brahic}. Thus our yet another future direction could be to combine these two ideas and possibly to include \cite{lazarev:extension} for studying general higher extensions and their integrations.

\vspace{3mm}

\noindent{\bf Notations:} Throughout the paper,
\emptycomment{we use $\huaE$ to
denote a 2-term complex of vector bundles
$\partial:E_{-1}\longrightarrow E_0$ and $E_\bullet$ is the
corresponding abelian Lie 2-groupoid (Example \ref{ex:abelian2g}).
$\huaE_n$ is a graded vector bundle $
\huaE_n:E_{-(n-1)}\oplus E_{-(n-2)}\oplus \cdots\oplus E_0. $  $\huaE_n[l]$ denotes the
$l$-shifted graded vector bundle, namely $(\huaE_n[l])_i=E_{l+i}$.
}
for a graded vector bundle $\huaE=\sum_qE_q$, $\huaE[l]$ denotes the
$l$-shifted graded vector bundle, namely $(\huaE[l])_i=E_{l+i}$.
 $(A,[\cdot,\cdot]_A,\rho_A)$
is a Lie algebroid and $d_A$ is the corresponding differential
defined by \eqref{eq:da}. We use $\huaG$ to denote a Lie groupoid
$G_1\rightrightarrows G_0$ and $\mathscr{G}$ to denote the Lie
groupoid $G_2\rightrightarrows G_1$ in a semistrict Lie 2-groupoid
$G_2\rightrightarrows G_1\rightrightarrows G_0$. $d$ is the usual de
Rham differential. We use $id$ to denote the identity map of vector
bundles and $id_M (resp. ~id_{G_0})$ is the identity map on the
manifold $M (resp. ~G_0)$. \vspace{3mm}

\noindent {\bf Acknowledgement}: We give our warmest thanks to David
Li-Bland, Zhang-Ju Liu, Rajan Mehta, Yvette Kosmann-Schwarzbach,
Weiwei Pan, Dimitry Roytenberg, Pavol \v{S}evera, Xiang Tang, and
Marco Zambon for very helpful discussions. The second author thanks
the invitation of the organizers to the conference ``Higher
structures in mathematics and physics'' in Z\"urich where this work
was initiated. In this conference she benefited a lot from various
discussions with, and a note shared by David Li-Bland.  She also
thanks Prof. Zhang-Ju Liu for his invitation to BICMR, where much of
the progress on this work was made.

\section{Preliminaries}

In this section, we recall important concepts that we will use in
our paper.

\subsection{NQ manifolds and Lie $n$-algebroids } 
  NQ manifolds   have their origin from the AKSZ and BV
quantization procedure \cite{aksz, schwarz:BV}. It is then made
popular in the area differential geometry due to the work of
Roytenberg and \v{S}evera \cite{royt,SWLett,s:funny}.

Recall that a degree $n$ {\bf N-manifold} $\huaM$ (for $n \in \Z_{\ge
  0} \sqcup{\infty}$) consists of a pair of a topological
space $M$ and a sheaf $C(\huaM)$ (viewed as the ring of functions of $\huaM$) over $M$ of graded commutative
algebras, locally isomorphic to $C^\infty(U)\otimes \Sym(\huaV^*)$, where
$U$ is an open set of $\R^m$ and $\huaV=\oplus_{i=-1}^{-n} V_i$ (and
when $n=0$, $\huaV=0$) is a
finite dimensional $\Z_{<0}$-graded vector space. Notice that it
equips
$M$ a structure of a smooth manifold and a degree 0 N-manifold is a
smooth manifold.

Then an {\bf NQ-manifold} is an N-manifold
$\huaM$ together with a degree
  $1$ vector field $Q$ satisfying $[Q,Q]=0$, i.e. a linear operator
  $Q$ on $C(\huaM)$ that raises the degree by one and  satisfying
  $Q^2=0$ and
  $$
Q(fg)=Q(f)g+(-1)^{|f|}fQ(g),\quad\forall~f,g\in C(\huaM).
  $$ Here the degree $|f|$ comes from the natural degree of the local
  model.
It is well known that a degree 1 NQ manifold corresponds to a Lie algebroid,
  thus we may think that
\begin{center}
  An NQ-manifold of degree $n$ corresponds to a Lie $n$-algebroid.
\end{center}

Some work in this direction appears in
\cite{Voronov:2010halgd}. Strictly speaking, a Lie $n$-algebroid gives
arise to an NQ-manifold only after a degree 1 shift, just as a Lie
algebroid $A$ corresponds to a degree 1 NQ manifold $A[1]$. To make
the shifting manifest, and to present a Lie $n$-algebroid in a way more
used to differential geometers, that is, to use the language of vector
bundles, we focus specifically on the split case. Moreover, it turns out that  semidirect products and
  extensions of Lie algebroids by   representations
up to homotopy provide some natural examples of these split NQ
manifolds.

Recall that a split degree $n$ N-manifold $\huaM$ is given by a
non-positively graded vector bundle $\huaA:=A_0 \oplus A_{-1} \oplus \cdots\oplus A_{-n+1}$ over the same base $M$, with the
function ring $C(\huaM)=C(\huaA[1])$ given by \eqref{eq:func-ring}. In
fact,  there is always a graded vector bundle $\huaA$ such
that $C(\huaM)\cong C(\huaA[1])$. However, the choice of $\huaA$ is not
canonical and is only unique up to an isomorphism. Nevertheless,  the space of
such choices is contractible.  Therefore, when we say that an
N-manifold is split, we actually equipped it with an extra
structure, namely a specific graded vector bundle $\huaA$ from which the
function ring $C(\huaM)$ comes. Then
$C(\huaA[1])$, equipped with the commutative product coming from
$C^\infty(M)$ and the symmetric product on $\Gamma(\Sym(\huaA[1])^*)$,
should be
viewed as  the Chevalley-Eilenberg complex of a certain higher
algebroid structure on $\huaA$  with brackets $l_i$'s and an anchor
$\rho: A_0\to TM $ such that  a degree $1$ homological vector field $Q$ can be
expressed by $\rho$ and  $l_i$'s:
\begin{equation} \label{eq:q-brackets}
\left\{\begin{array}{rll}
  Q(f) &=&\rho^*(df), \\
Q(\xi^k)(X^{i_1}_1, \dots, X^{i_j}_j) &=&- \langle \xi^k, l_j(X^{i_1}_1, \dots,
X^{i_j}_j)\rangle, \quad j\neq 2,\\
Q(\xi^k)(X_1^0, X_2^k)&=& \rho(X^0_1) \langle \xi^k, X^k_2 \rangle -
\langle \xi^k, l_2(X_1^0, X_2^k) \rangle, \quad k\neq0,\\
Q(\xi^0)(X_1^0, X_2^0)&=& \rho(X^0_1) \langle \xi^0, X^0_2 \rangle -\rho(X^0_2) \langle \xi^0, X^0_1 \rangle-
\langle \xi^0, l_2(X_1^0, X_2^0) \rangle,
\end{array}\right.
\end{equation}
for all $f\in C^\infty(M), \xi^k\in\Gamma(A_{-k}^*), X^{i_p}_p\in \Gamma(A_{-i_p}),~i_1+\cdots+i_j=k+2-j$. That is, in general, $Q$ is the dual of $l_i$'s, except for
$l_2$, where terms of derivation involving $\rho$ appear. See \cite[Lemma 5]{malte-thesis} and \cite[Theorem 2]{BP} for more details.

 \emptycomment{For the
first several $l_i$'s we have,
\begin{eqnarray*}
\begin{split}
Q(\xi^0)( X^1) &=-\langle\xi^0,l_1(X^1)\rangle,\\
Q(\xi^0)(X^0, Y^0) &=\rho(X^0)\langle\xi_0,Y^0\rangle-\rho(Y^0)\langle\xi_0,X^0\rangle-\langle\xi^0,l_2(X^0,Y^0)\rangle,\\
  Q(\xi^1)(X^0, Y^0, Z^0) &=-\langle\xi^1,l_3(X^0,Y^0,Z^0)\rangle,\\
   Q(\xi^1)(X^0,  X^1) &=\rho(X^0)\langle\xi_1,X^1\rangle-\langle\xi^1,l_2(X^0,X^1)\rangle,
\end{split}
\end{eqnarray*}
for all   $X^0,Y^0,Z^0\in\Gamma(A_{0})$ and $X^1\in\Gamma(A_{-1})$. See \cite[Theorem 2]{BP} for details.
}

As in the
case of $L_\infty$-algebras, the condition that $(C(\huaA[1]), Q)$ is a differential
graded commutative algebra, namely $Q^2=0$,  is equivalent to all
the axioms that $l_i$ and $\rho$ should satisfy. Finally, we
summarize this equivalent viewpoint of a split NQ manifold with the
following definition:

\begin{defi}[split Lie $n$-algebroid]\label{def:n-algd}
  A split Lie $n$-algebroid is a non-positively graded vector bundle $\huaA=A_0\oplus A_{-1}\oplus\cdots\oplus
  A_{-n+1}$ over a manifold $M$ equipped with a bundle map $\rho:A_0\longrightarrow
  TM$ (called the anchor), and $n+1$ many brackets $l_i:\Gamma(\wedge^i\huaA)\longrightarrow \Gamma(\huaA)$ with
  degree $2-i$ for $1\le i \le n+1$,  such that
\begin{itemize}
  \item[\rm 1.]
\begin{equation} \label{eq:l-infty-brackets}
\sum_{i+j=k+1}(-1)^{i(j-1)}\sum_{\sigma \in Sh^{-1}_{i, k-i}}\sgn(\sigma)\Ksgn(\sigma)l_j(l_i(X_{\sigma(1)},\cdots,X_{\sigma(i)}),X_{\sigma(i+1)},\cdots,X_{\sigma(k)})=0,
\end{equation}
where the summation is taken over all $(i,k-i)$-unshuffles $ Sh^{-1}_{i, k-i}$ with
$i\geq1$ and ``$\Ksgn(\sigma)$'' is the Koszul sign for a
permutation $\sigma\in S_k$, i.e. $$ X_1\wedge  \cdots\wedge
X_k=\Ksgn(\sigma)X_{\sigma(1)}\wedge  \cdots\wedge
X_{\sigma(k)}.
$$

\item[\rm 2.]  $l_2$ satisfies the Leibniz rule with respect to $\rho$:
$$l_2(X^0,fX)=fl_2(X^0,X)+\rho(X^0)(f)X,\quad\forall~X^0\in\Gamma(A_0),
f\in C^\infty(M), X\in\Gamma(\huaA).$$

\item[\rm 3.] For $i\neq 2$,  $l_i$'s are $ C^\infty(M)$-linear.
\end{itemize}
\end{defi}
\begin{rmk}\label{rmk:no-dual-version}
We see that $Q$ made up by $l_i$'s as in \eqref{eq:q-brackets} is a
degree $1$ homological vector field on $C(\huaA[1])$.  Thus, a split Lie $n$-algebroid corresponds to
a split degree $n$ NQ manifold. The equivalence of the category of split NQ
manifolds and the category of split higher Lie algebroids (in finite
dimensional case), known
to experts in the domain for a long time, is later
rigorously proven in \cite{BP}.

We also notice that one of the advantages of the concept of a split Lie $n$-algebroid is that it does not involve
the dual of a vector bundle. Thus, it is convenient in the infinite
dimensional case.

Finally, we remark that the $2$-term case of split higher Lie algebroids
is further much studied in
\cite{Jot}.

\emptycomment{Another way to avoid dual is to consider the
following graded vector bundle over $M$
\[ Sym(\huaA[1]) =  [A_0] \oplus \big[ \wedge^2 A_0 \oplus
A_{-1}\big] \oplus \big[ \wedge^3 A_0 \oplus A_0 \otimes A_{-1} \oplus
A_{-2}\big] \oplus \dots,  \]
with a coalgebra structure $\Delta$ on the sections
\[\Delta (x_1 \wedge \dots \wedge x_k ) = \sum_i \sum_{\sigma\in
  Sh^{-1}_{i, k-i}} \Ksgn(\sigma) (x_{\sigma(1)} \wedge \dots \wedge
x_{\sigma(i)} ) \otimes (x_{\sigma(i+1)} \wedge \dots \wedge
x_{\sigma(k)}).
\] Here we abuse the notation $\wedge$ to denote the product in
the symmetric algebra (as it denotes also the product in the
exterior algebra). Then $\rho$ and the $l_i$'s combine to form a degree
1 codifferential $d$ on $Sym(\huaA[1])$. That is, $\Delta \circ d =
(d \otimes 1 + 1\otimes d) \circ \Delta $ and $d^2=0$.}
\end{rmk}
\begin{rmk} \label{rmk:sections}
We see that the brackets $l_i$'s make the space of smooth
sections $\Gamma(A_0) \oplus \Gamma(A_{-1})\oplus \cdots \oplus
\Gamma(A_{-n+1})$ of a split Lie $n$-algebroid $\huaA$ into a (infinite
dimensional) Lie $n$-algebra.
\end{rmk}

\begin{defi}\label{def:morphism}
A split Lie $n$-algebroid morphism (resp. isomorphism) $\huaA \to
\huaA'$ (possibly over different base manifold) is
a graded vector
bundle morphism $f: \Sym(\huaA[1]) \to \Sym(\huaA'[1])$
such that the induced map $f^*:
C(\huaA'[1]) \to C(\huaA[1])$ is a morphism (resp. isomorphism) of NQ
manifolds.
\end{defi}

\begin{remark} \label{remark:isom-lie-n}
First of all, when $n=1$ this definition coincides with the usual definition of
a Lie algebroid morphism (see e.g. \cite[Section 4.3]{MK2}) in terms of vector
bundles, anchors and brackets, which is more involved than this
one-sentence definition.
For morphisms of
split Lie $n$-algebroids in terms of vector bundles, anchors and
brackets, please see \cite[Section 4.1]{BP}.
In the case when the $\huaA$ and $\huaA'$ share the same
base and the base morphism is an isomorphism, a graded vector bundle
morphism $f$ is a split Lie $n$-algebroid morphism  if and only if $f$
preserves the anchors and $f$ induces an $L_\infty$-morphism on the
sections of $\huaA$ and $\huaA'$.   Notice that this implies
that $f$ preserves the brackets only in an $L_\infty$-fashion. For
example, when $n=2$, we have
\begin{eqnarray}
f_0(l_2(X^0, Y^0))- l'_2(f_0(X^0),
f_0(Y^0))&=&  l_1'(f_2(X^0, Y^0))\label{eq:l2l'2},\\
f_1(l_2(X^0, X^1))-l'_2(f_0(X^0), f_1(X^1))&=& f_2(X^0, l_1(X^1))
\label{eq:f2l2}, \\
\nonumber l'_3(f_0(X^0), f_0(Y^0),
f_0(Z^0))-f_1(l_3(X^0, Y^0, Z^0))
&=&l'_2(f_2(X^0, Y^0), f_0(Z^0))\\
&&+ f_2(l_2(X^0, Y^0), Z^0)+c.p.(X^0,Y^0,Z^0).
\label{eq:l3l'3}
\end{eqnarray}
where $X^0, Y^0, Z^0\in\Gamma(A_0),~X^1\in\Gamma(A_{-1})$ and
$c.p.(X^0,Y^0,Z^0)$ means cyclic permutations of $X^0,Y^0,Z^0$.

Notice that in some special situation, $f$ might be induced from a vector
bundle morphism $\bar{f}: \huaA\to\huaA'$, that is,
$f=\Sym(\bar{f})$. In this case, $f$ preserves the bracket strictly,
and we call such $f$ a {\bf strict
  morphism}.   If a split Lie $n$-algebroid morphism $f$
induces an isomorphism on the underlying graded vector bundle of
$\huaA$ and $\huaA'$,  then it is an isomorphism.
\end{remark}

\begin{ep}{\rm \label{ep:lower-algd}
First of all, we see that a Lie algebroid (namely a Lie $1$-algebroid)
$A$ is always a split Lie $1$-algebroid. Moreover, a Lie algebroid $A$ can be viewed as a split Lie
$n$-algebroid with $A_0=A$, all the other $A_i=0$, the same anchor
and $l_2$, and all the other higher brackets equal to 0. In fact,
a split Lie $n$-algebroid $\huaA$ can be viewed as a split Lie
$(n+1)$-algebroid in this way. That is, $\huaA$ is a split Lie
$(n+1)$-algebroid with $A_{-n}=0$, $l_{n+2}=0$, and all the rest
are the same.
}
\end{ep}

\begin{ep}{\rm \label{ep:ab-algd}
Given a complex of vector bundles
$\huaE_n:E_{-(n-1)}\stackrel{\partial}{\lon}
E_{-(n-2)}\stackrel{\partial}{\lon} \cdots\stackrel{\partial}{\lon}
 E_0$,
 it can be viewed as
a split Lie $n$-algebroid with $l_1=\partial$, the remaining brackets
$l_i=0$ for $i\ge 2$, and the anchor $\rho=0$. We call such a split Lie $n$-algebroid {\bf abelian}. There are similar constructions
on the groupoid level as we point out in the case when $n=2$ in
Example \ref{ex:abelian2g}. The integration and differentiation
between these abelian higher algebroids and groupoids are given by
Dold-Kan correspondence (see for example, \cite[Example 7]{LS} for a
detailed explanation in this direction).}
\end{ep}

\subsection{Representations up to homotopy of higher Lie
algebroids}\label{rep}

\begin{defi}
Given a degree $n$ NQ manifold $(\huaM, Q) $ over $M$,
a representation up to homotopy of $(\huaM, Q)$ is a graded vector
bundle
\[ \huaE =E_{i-m}\oplus \cdots \oplus E_{i-1}\oplus
  E_i, \]
over $M$, together with a degree $1$ differential \footnote{This means that
  $D$ is an $\R$-linear operator increasing the total degree by 1 and $D^2=0$.}
  $
  D:  C(\huaM)\otimes \Gamma \big(\huaE \big)
  \longrightarrow C(\huaM)\otimes \Gamma \big( \huaE \big)
  $
satisfying the graded derivation rule:
  \begin{equation}\label{eq:gradedderivation}
    D(\omega\eta)=(Q\omega) \eta+(-1)^k\omega D(\eta),\quad
    \forall~\omega\in C^k(\huaM),\quad \forall
    \eta\in  C(\huaM) \otimes \Gamma\big(\huaE \big).
  \end{equation}
Here $C(\huaM)\otimes \Gamma \big( \huaE \big)$ is equipped with a
natural structure of a graded $C(\huaM)$-module with degree
$|\omega \otimes \epsilon|= k+j$, where $\omega\in C^k(\huaM)$ and
$\epsilon \in E_j$. Given a split Lie $n$-algebroid $(\huaA, \{l_i\}_{i=1}^{n+1},
\rho)$, it corresponds to a degree $n$ NQ manifold $(\huaA[1], Q)$. A
representation up to homotopy of $(\huaA, \{l_i\}_{i=1}^{n+1}, \rho)$ is defined as such a representation up to homotopy of $(\huaA[1], Q)$.
\end{defi}

\begin{rmk}
When $n=1$, we have a Lie ($1$-)algebroid. Thus this definition gives
rise to that of a representation
up to homotopy of a Lie algebroid given in
\cite{abad-crainic:rep-homotopy}.
\end{rmk}
\emptycomment{
\begin{defi}{\rm\cite{abad-crainic:rep-homotopy}}
Given a Lie algebroid $(A,[\cdot,\cdot]_A,\rho_A)$ over base manifold
$M$,   a representation up to homotopy of $A$ is a graded vector
bundle
\[ \huaE_n =E_{-(n-1)}\oplus E_{-(n-2)}\oplus \cdots\oplus
  E_0, \]
over $M$, together with a degree 1 differential \footnote{This means that
  $D$ $\R$-linear operator increasing the total degree by 1 and $D^2=0$.}
  $
  D:\Gamma \big( C( A[1])\otimes \huaE_n \big)
  \longrightarrow \Gamma \big( C(A[1])\otimes \huaE_n \big)
  $
satisfying the graded derivation rule:
  \begin{equation}\label{eq:gradedderivation}
    D(\omega\eta)=(d_A\omega) \eta+(-1)^k\omega D(\eta),\quad
    \forall~\omega\in\Gamma\big(\Sym^kA[1]^*\big),\quad \forall
    \eta\in \Gamma\big( C(A[1])\otimes \huaE_n \big).
  \end{equation}
\end{defi}
}

\emptycomment{
\comment{we might take away all the dual stuff.}
\begin{rmk}
To avoid taking duals in the above definition, an equivalent method
is to define co-representations: A {\em co-representation up to homotopy}
on a graded vector space $\huaF$ is a codifferential $D$ of
$\Gamma \big( \Sym (A  [1]) \otimes \huaF \big), \Delta)$ with $\Delta$ similarly defined as the
co-product of $Sym \huaA[1]$. For more details, see
\cite{malte-thesis}.
\end{rmk}
}
We denote  a representation up to homotopy of $(\huaM, Q)$ by $(\huaE,D)$,
and the set of all such representations of $(\huaM, Q)$ by
$Rep^\infty(\huaM, Q)$. We also denote the set of all such
representations of a split Lie $n$-algebroid $(\huaA, \{l_i\}_{i=1}^{n+1}, \rho)$ by
$Rep^\infty(\huaA, \{l_i\}_{i=1}^{n+1}, \rho)$.
Such a representation is the
Lie algebroid version of Lada-Markl's $L_\infty$-module
\cite{LadaMartin}.

We now write down the explicit formula for a 2-term representation up
to homotopy for a Lie algebroid $A$. This is the case we need to use
for exact Courant algebroids.

\begin{ep} {\rm\cite{abad-crainic:rep-homotopy}} \label{pro:rep-homotopy}
{\rm
Given a Lie algebroid $A$, a $2$-term representation up to homotopy
$(E_{-1} \oplus E_0,D)$ is made up by a differential $\partial:
E_{-1} \to E_0$ given by the component of $D$ restricted on $E_{-i}$'s $(i=0,1)$,
an $A$-connection $\nabla=(\nabla^{-1},\nabla^0)$ on $E_{-1} \stackrel{\partial}{\longrightarrow} E_0$
 given by the components of $D$ restricted on $\Gamma(E_{-i})\longrightarrow \Gamma(A^*\otimes E_{-i})$, and an element $\omega \in\Gamma(\wedge^2A^*\otimes \Hom(E_0,E_{-1}))$ satisfying
 $$
 R_{\nabla^{-1} }(X,Y)(m)=-\omega(X,Y)(\partial m),\quad R_{\nabla^0 }(X,Y)(u)=-\partial(\omega(X,Y)(u)),
 $$
 for all $  X,Y\in\Gamma(A), ~u\in\Gamma(E_0),~m\in\Gamma(E_{-1}),$ and
 $$
  \overline{d}_\nabla\omega=0.
 $$
 Here   $\nabla^{-1}$ and $\nabla^0$ are $A$-connections on $E_{-1}$ and $E_0$ satisfying $\partial\nabla^{-1}_Xm=\nabla^0_X(\partial m)$ for all $X\in\Gamma(A)$ and $m\in\Gamma(E_{-1})$,  $R_{\nabla^{-i}}$ is the curvature of $\nabla^{-i}$ and $\overline{d}_\nabla\omega\in\Gamma(\wedge^3A^*\otimes \Hom(E_0,E_{-1}))$ is given by
 $$
\overline{ d}_\nabla\omega(X,Y,Z)=[\nabla_X,\omega(Y,Z)]-\omega([X,Y]_A,Z)+c.p.(X,Y,Z), \quad\forall X,Y,Z\in\Gamma(A),
 $$
 in which $[\nabla_X,\omega(Y,Z)]\in\Gamma(\Hom(E_0,E_{-1})$ is given by
 $$
 [\nabla_X,\omega(Y,Z)](u)=\nabla^{-1}_X(\omega(Y,Z)(u))-\omega(Y,Z)(\nabla^0_Xu),\quad \forall u\in\Gamma(E_0).
 $$
  The correspondence is given by
 $$D=\partial+d_\nabla+\omega,$$
 where $d_\nabla:\Gamma(\wedge^kA^*\otimes E_{-i})\longrightarrow \Gamma(\wedge^{k+1}A^*\otimes E_{-i})$ is the usual operator given by the Koszul formula, $\partial $ is viewed as the map from $\Gamma(\wedge^kA^*\otimes E_{-1})$ to $\Gamma(\wedge^{k}A^*\otimes E_0)$ via
 $$
 \partial (\varpi)(X_1,\cdots X_k)=(-1)^k\partial \big( \phi(X_1,\cdots X_k)\big),\quad \forall \varpi\in\Gamma(\wedge^kA^*\otimes E_{-1}), X_i\in\Gamma(A),
 $$
  and $\omega$ is viewed as a map from $\Gamma(\wedge^kA^*\otimes E_0)$ to $\Gamma(\wedge^{k+2}A^*\otimes E_{-1})$ via
 \[
 \omega(\varpi)(X_1,\cdots,X_{k+2})=\sum_{i<j}-(-1)^{k+i+j} \omega(X_i,X_j)\varphi(X_1,\cdots,\widehat{X_i},\cdots,\widehat{ X_j},\cdots, X_{k+2}),
 \]
 Thus, we also denote a 2-term representation up to homotopy by $(E_{-1}
\xrightarrow{\partial} E_0, \nabla, \omega)$.
 }
\end{ep}

\emptycomment{
\begin{pro}{\rm\cite{abad-crainic:rep-homotopy}} \label{pro:rep-homotopy}
There is a one-to-one correspondence between representation up to
homotopy $(\huaE_n,D)$ of $A$ and graded vector bundles $\huaE_n$
over $M$ endowed with
\begin{itemize}
  \item[\rm 1.] A degree $1$ operator $\partial $ on $\huaE_n$ making
  $(\huaE_n,\partial)$ a complex;
  \item[\rm 2.] An $A$-connection $\nabla$ on $(\huaE_n,\partial)$;
  \item[\rm 3.] An $\End(\huaE_n)$ valued $2$-form $\omega_2$ of total
  degree $1$, i.e. $\omega_2\in\Omega^2(A,\End^{-1}(\huaE_n))$
  satisfying $\partial\omega_2+R_\nabla=0$, where $R_\nabla$ is the curvature of $\nabla$.
  \item[\rm 4.] For each $i>2$ an $\End(\huaE_n)$-valued $i$-form
  $\omega_i$ of total degree $1$, i.e. $\omega_i\in\Omega^i(A,\End^{1-i}(\huaE_n))$
  satisfying
  $$
\partial
\omega_i+d_\nabla\omega_{i-1}+\omega_2\circ\omega_{i-2}+\cdots+\omega_{i-2}\circ\omega_2=0.
  $$
\end{itemize}
The correspondence is characterized by
$$
D(\eta)=\partial
\eta+d_\nabla\eta+\omega_2\circ\eta+\omega_3\circ\eta+\cdots.
$$
\end{pro}
We also write $D=\partial+d_\nabla+\omega_2+\cdots.$}

Let $(\huaM, Q)$ be a degree $n$ NQ manifold. There are many operations in $Rep^\infty(\huaM, Q)$.
First of all, it is easy to see that if $(\huaE, D)\in Rep^\infty
(\huaM, Q)$, then any {\bf shift} $\huaE[r]$ is again in $Rep^\infty(\huaM, Q)$ with the same $D$.
We can also take the {\bf dual} of
$(\huaE,D)$. Consider the dual  complex $\huaE^*$. Then there is an operator
$D^*:C(\huaM) \otimes\Gamma(\huaE^*) \longrightarrow C(\huaM) \otimes\Gamma(\huaE^*) $ uniquely
determined by the condition
\begin{equation}
  Q(\eta\wedge\eta^\prime)=D^*(\eta)\wedge\eta^\prime+(-1)^{|\eta|}\eta\wedge
  D(\eta^\prime),\quad\forall~\eta\in C(\huaM)\otimes
  \Gamma(\huaE^*),\eta^\prime\in C(\huaM) \otimes\Gamma(\huaE),
\end{equation}
where $\wedge$ is the operation
$[C(\huaM)\otimes
  \Gamma(\huaE^*)] \otimes [ C(\huaM) \otimes\Gamma(\huaE)]
  \longrightarrow C(\huaM)$
induced by the natural pairing between $\huaE^*$ and $\huaE$. Then
$(\huaE^*,D^*)$ is a representation up to homotopy of $(\huaM, Q)$. \emptycomment{In terms
of the components of $D$, if $D=\partial +\nabla+\sum_{i\geq2}\omega_i$,
then we find that $D^*=\partial^*+\nabla^*+\sum_{i\geq2}\omega_i^*$,
where $\nabla^*$ is the connection dual to $\nabla$ and,
\begin{eqnarray}
  \partial^*\eta_k=(-1)^{k+1}\eta_k\circ\partial,  \quad
  \omega_p^*(X_1,\cdots,X_p)(\eta_k)=-(-1)^{k(p+1)}\eta_k\circ\omega_p(X_1,\cdots,X_p),
\end{eqnarray}
for any $
  \eta_k\in E_{-k}^*$ and $X_1,\cdots,X_p\in\Gamma(A).$}

For two representations up to homotopy, $(\huaE, D^{\huaE})$ and
$(\huaF, D^{\huaF})$ of $(\huaM, Q)$, one can also take their {\bf tensor
product}. Consider the operator $D$ corresponding to $\huaE\otimes
\huaF$, which is uniquely determined by the condition
\begin{equation} \label{eq:D-tensor}
  D(\eta_1\wedge\eta_2)=D^{\huaE}(\eta_1)\wedge\eta_2+(-1)^{|\eta_1|}\eta_1\wedge
  D^{\huaF}(\eta_2),\quad\forall~\eta_1\in C(\huaM)
  \otimes\Gamma(\huaE),~\eta_2\in  C(\huaM) \otimes\Gamma(\huaF),
\end{equation}
where $\wedge$ is the natural operation $[C(\huaM)
  \otimes\Gamma(\huaE)] \otimes [C(\huaM)\otimes \Gamma(\huaF)]\to
  C(\huaM) \otimes \Gamma (\huaE \otimes \huaF)$.
Then $(\huaE \otimes \huaF,  D)$ is a representation up to
homotopy of $(\huaM, Q)$. \emptycomment{In term of components, if
$D^{\huaE_n}=\partial^{\huaE_n}+\nabla^{\huaE_n}+\sum_{{i\geq2}}\omega_i^{\huaE_n}$
and
$D^{\huaF_m}=\partial^{\huaF_m}+\nabla^{\huaF_m}+\sum_{{i\geq2}}\omega_i^{\huaF_m}$,
then $D=\partial +d_\nabla+\sum_{i\geq2}\omega_i$, where
\begin{itemize}
  \item[\rm 1.] $\partial$ is the tensor product of $\partial^{\huaE_n}$ and
  $\partial^{\huaF_m}$: $\partial(u\otimes v)=\partial^{\huaE_n} u\otimes v+(-1)^{|u|}u\otimes\partial^{\huaF_m} v$;
  \item[\rm 2.] $\nabla$ is the tensor product of $\nabla^{\huaE_n}$ and
  $\nabla^{\huaF_m}$:
  $$
\nabla_X(u\otimes v)=\nabla_X^{\huaE_n} u\otimes
v+(-1)^{|u|}u\otimes\nabla_X^{\huaF_m} v,\quad\forall~X\in\Gamma(A),
u\in \Gamma({\huaE_n}),v\in\Gamma( {\huaF_m});
  $$
  \item[\rm 3.] $\omega_p=\omega_p^{\huaE_n}\otimes id+id\otimes\omega_p^{\huaF_m}.$
\end{itemize}
}

Similarly, if $\huaE \in Rep^\infty(\huaM, Q)$, then the symmetric vector
bundle complex $\Sym(\huaE)$ and the exterior vector bundle complex $\Lambda
(\huaE)$ are again in $Rep^\infty(\huaM, Q)$.

\subsection{Courant algebroids}\label{sec:courant}

 A Courant algebroid is a vector bundle $E\longrightarrow M$ equipped
with a nondegenerate symmetric bilinear form $\pair{\cdot,\cdot}$ on
the bundle, an antisymmetric bracket $\Courant{\cdot,\cdot}$ on the
section space $\Gamma(E)$ and a bundle map $\rho:E\longrightarrow
TM$ such that a set of axioms are satisfied (see \cite[Definition 2.1]{lwx}).

We pay special attention to  the {\bf exact  Courant algebroid}
$(\huaT=TM\oplus
T^*M,\pair{\cdot,\cdot},\Courant{\cdot,\cdot},\rho)$ associated to a
manifold $M$ with \v{S}evera class $[H]$ with $H \in \Omega_{\rm cl}^3(M)$.
Here the anchor $\rho:\huaT\longrightarrow TM$ is the projection.
The canonical pairing $\pair{\cdot,\cdot}$ is given by
\begin{equation}\label{eqn:pair}
\pair{X+\xi,Y+\eta}=\frac{1}{2}\big(\xi(Y)+\eta(X)\big),\quad\forall
~X,Y\in\frkX(M),~\xi,\eta\in\Omega^1(M).
\end{equation}
The antisymmetric bracket $\Courant{\cdot,\cdot}$ is given by
\begin{equation}\label{eqn:bracket}
\Courant{X+\xi,Y+\eta}\triangleq[X,Y]+L_X\eta-L_Y\xi+\frac{1}{2}d(\xi(Y)-\eta(X))+i_{X\wedge
Y}H.
\end{equation}
It is not a Lie bracket, but we have
\begin{equation}\label{eqn:jacobi}
\Courant{\Courant{e_1,e_2},e_3}+\Courant{\Courant{e_2, e_3}, e_1} +
\Courant{\Courant{e_3, e_1}, e_2}=d T(e_1,e_2,e_3),\quad
\forall~e_1,e_2,e_3\in\Gamma(\huaT),
\end{equation}
where $T(e_1,e_2,e_3) $ is given by
\begin{equation}\label{T}
T(e_1,e_2,e_3)=\frac{1}{3}(\pair{\Courant{e_1,e_2},e_3}+\pair{\Courant{e_2,e_3},e_1}+\pair{\Courant{e_3,e_1},e_2}).
\end{equation}
As shown in \cite{royt}, there is a one-to-one correspondence between degree
2 NQ manifolds with degree 2 symplectic forms and Courant
algebroids. The degree 2 NQ manifold corresponding to $(\huaT=TM\oplus
T^*M,\pair{\cdot,\cdot},\Courant{\cdot,\cdot},\rho)$  is
$T^*[2]T[1]M$, with degree 2 symplectic form the canonical
symplectic form on a cotangent bundle. In local coordinates, the
symplectic form is $dp_i  dq^i+d\xi^i
d\theta_i$ and the homological vector field  is
 \begin{equation}\label{eq:Q}
Q=\xi^i\frac{\partial}{\partial q^i}+p_i\frac{\partial}{\partial
\theta_i}+\frac{1}{6}\frac{\partial \phi_{ijk}(q)}{\partial
q^l}\xi^i\xi^j\xi^k\frac{\partial}{\partial p_l}-\frac{1}{6}
\phi_{ijk}(q)\xi^i\xi^j\frac{\partial}{\partial \theta_k}.
\end{equation}
This is exactly the
Hamiltonian vector field of
$\xi^ip_i-\frac{1}{6}\phi_{ijk}\xi^i\xi^j\xi^k$, where $\phi_{ijk}$ is defined in terms of the 3-form $H=\frac{1}{6}\phi_{ijk}\xi^i\xi^j\xi^k$.  Here $q^i$'s are coordinates on $M$, $\xi^i=dq^i$
are cotangent vectors, thus coordinates on the fibre of $T[1]M$, moreover $
p_i=\frac{\partial}{\partial q^i}$ and
$\theta_i=\frac{\partial}{\partial \xi^i}$ are coordinates on the
cotangent fibre. The degrees of  $q^i,\xi^i,p_i,\theta_i$ are $0,1,2,1$
respectively. If $H=0$, such a Courant algebroid is a {\bf standard
  Courant algebroid}.

\section{Semidirect products for  representations up to homotopy}

Let $(\huaE=E_0\oplus E_{-1} \oplus \dots \oplus E_{-m+1}, D)$ be a representation up to homotopy of a split Lie $n$-algebroid
$(\huaA=A_0\oplus A_{-1} \oplus \dots \oplus A_{-n+1}, \{l_i\}_{i=1}^{n+1}, \rho)$. Let $\widetilde{D}$ be the
induced representation up to homotopy of $(\huaA, \{l_i\}_{i=1}^{n+1}, \rho)$ on
$\Sym((\huaE [1])^*) $. We may take their semidirect product and obtain
again a higher split Lie $n$-algebroid. We now make this explicit in the
following statement.

\begin{pdef}\label{lem:ruthLie2a}
With the above notations,  $\widetilde{D}$ is a degree $1$ homological
vector field on $C((\huaA\oplus \huaE)[1])$, and this makes the graded vector bundle
 \[\huaA \oplus \huaE= [A_0\oplus E_0] \oplus [A_{-1} \oplus E_{-1}]
 \oplus \dots \] a split Lie $k$-algebroid, where $k=\max(n, m)$. We call this Lie
 $k$-algebroid the {\bf semidirect-product} of the split Lie
$n$-algebroid $(\huaA, l_i, \rho)$ with its representation up to homotopy $(\huaE, D)$, and denote it by
$\huaA\ltimes_{D} \huaE$(or simply $\huaA\ltimes \huaE$ if there is
no confusion in the context).
\end{pdef}
\pf  The representation up to homotopy $\widetilde{D}$  of $(\huaA,
\{l_i\}_{i=1}^{n+1}, \rho)$ on
$\Sym(\huaE[1]^*)$, is a degree 1 differential  on
\[C(\huaA[1])\otimes \Gamma( \Sym (\huaE[1])^*) = C^\infty (M) \oplus
\Gamma(\Sym (\huaA[1])^*
) \otimes \Gamma(\Sym (\huaE[1])^*) = C ((\huaA\oplus
\huaE)[1]),
\] satisfying \eqref{eq:gradedderivation}. At this moment, $\widetilde{D}$ is only an $\R$-linear differential
on the $C(\huaA[1])$ module $C(\huaA[1]) \otimes
\Gamma(\Sym(\huaE[1]^*))$.  But notice that the graded commutative algebra
structure on $C ((\huaA\oplus
\huaE)[1])$ comes from the product on $C(\huaA[1])$ and the
symmetric product
on $\Gamma(\Sym(\huaE[1]^*)$. Thus $\widetilde{D}$ becomes a
graded derivation with respect to the commutative product on $C ((\huaA\oplus
\huaE)[1])$ thanks to the extending rule
\eqref{eq:D-tensor}. Therefore $ \widetilde{D}$ is a degree 1 homological vector
field on $\huaA\oplus \huaE$, and this makes $\huaA\oplus \huaE$ a
split Lie $k$-algebroid.
\qed
\begin{rmk}
This semidirect product can be made also for a general Lie $n$-algebroid
$(\huaM, Q)$. Applying the same procedure, we obtain a Lie
$k$-algebroid with function ring $C(\huaM) \otimes
\Gamma(\Sym(\huaE [1])^*)$. We emphasis the split version here because it is
especially nice that the semidirect product remains split with a simple
and clear formula.

Nevertheless, for semidirect products, the
representation up to homotopy $\huaE$ must start from degree $0$. Otherwise,
for the result to be a higher Lie algebroid, namely the function ring to be
focused on the non-negative part, we will need to shift $\huaE$.

We will give a more conceptual explanation in
Section \ref{sec:ext} in the $2$-term case, where a semidirect product corresponds to an extension
with the trivial extension class in $H^2(A, \huaE)$.
\end{rmk}

\emptycomment{
\begin{rmk} Using co-representations, we can completely avoid taking duals in
above construction. Given a co-representation up to homotopy
$\huaF$, the induced co-representation $\tilde{D}$ on $Sym(A[1])
\otimes Sym(\huaF[1]) =Sym((A\oplus \huaF)[1])$ naturally gives a
Lie $n$-algebroid structure on $A \oplus \huaF$.
\end{rmk}
}

Notice that a Lie algebroid $A$ is always
split. Thus, the semidirect product of a Lie algebroid with its
representation up to homotopy is always
split. This provides a series of examples of interesting split higher
Lie algebroids. Now for the other direction, when will a split Lie
$n$-algebroid be a semidirect product of a Lie algebroid with a
representation up to homotopy? In the following statement, we give a
complete criteria.

\begin{pro}\label{thm:Qreputh}
Given a Lie algebroid $A$ over $M$, let $\huaE =E_{0}\oplus E_{-1} \oplus \dots \oplus E_{-1+n} $ be a graded vector bundle
over $M$. Suppose that there is a split Lie $n$-algebroid structure on
$(A\oplus E_0)\oplus E_{-1} \oplus \dots\oplus E_{-n+1}$  given
by a degree $1$ homological vector field $Q$,
\begin{equation} \label{eq:Q-semidirect-product}
\begin{split}
 &\CWM\xrightarrow{Q} \Gamma((A\oplus E_0)^*)\xrightarrow{Q} \Gamma( \wedge^2(A\oplus E_0)^*)\oplus
\Gamma(E_{-1}^*)\\
&\xrightarrow{Q}\Gamma( \wedge^3(A\oplus E_0)^*)\oplus
\Gamma((A\oplus E_0)^*)\otimes\Gamma(E_{-1}^*) \oplus
\Gamma(E^*_{-2})\xrightarrow{Q}\cdots.
\end{split}
\end{equation}
Then this Lie $n$-algebroid  is a semidirect product
$A\ltimes_D \huaE$, for a certain representation up to homotopy $D$,  if and only if the following conditions are
satisfied:
\begin{itemize}
  \item[\rm (a)] The restriction of $Q$ on $C^\infty(M) \oplus \Gamma(\wedge^\bullet A^*)$ is exactly given by $d_A$.

\item[\rm (b)] The image of $Q$ on $\Gamma(E^*_i)$ satisfies, 
\[Q(\Gamma(E_0^*))\subset\Gamma(E_{-1}^*)\oplus
\Gamma(A^*\otimes E^*_0),\]
\[Q(\Gamma(E_{-1}^*))\subset \Gamma(E_{-2}^*)\oplus \Gamma(A^*\otimes E_{-1}^*)\oplus
\Gamma(\wedge^2A^*\otimes E^*_0),\]
\dots
\[Q(\Gamma(E^*_{-n+2})) \subset \Gamma(E^*_{-n+1}) \oplus \Gamma(A^*
\otimes E_{-n+2}^*) \oplus \dots \oplus \Gamma(\wedge^{n-1} A^* \otimes
E^*_0) , \]
\[Q(\Gamma(E^*_{-n+1})) \subset \Gamma(A^* \otimes E_{-n+1}^* ) \oplus
\Gamma(\wedge^2 A^* \otimes E_{-n+2}^*) \oplus
\dots \oplus \Gamma(\wedge^{n} A^* \otimes E^*_0) .\]
\end{itemize} In this case, $D=Q|_{C(A[1])\otimes \Gamma(\huaE)}$.
\end{pro}
\pf From Conditions (a) and
(b), we see that $Q(\Gamma(\wedge^\bullet A^*)) $ does not contain any terms in
$E_\bullet$, and
\[ Q(\Gamma(E_i^*) )\cap
\big( \Gamma(\wedge^\bullet A^*) \oplus \Gamma(\wedge^{\ge 2}
E_\bullet^*) \big) =0. \] Such a vector field $Q$ maps
$C(\huaA[1])\otimes \Gamma(\huaE)$ to itself. The fact that $Q$ is a derivation satisfying
$Q^2=0$ implies that  $(\huaE,Q|_{C(A[1])\otimes \Gamma(\huaE)}) \in
Rep^\infty (A)$. It is not hard to see that $Q$ is an extension of
$Q|_{C(A[1])\otimes \Gamma(\huaE)}$ by graded Leibniz rule. Thus, the Lie
$n$-algebroid given by \eqref{eq:Q-semidirect-product} is exactly $A \ltimes
\huaE $. We leave the other direction  to the reader as an
exercise.
 \qed

\begin{rmk}\label{rmk:dual-semidirect-product}
Suppose that the Lie $n$-algebroid structure corresponding to $Q$ is
given by $\rho$ and $ \{l_i\}_{i=1}^{n+1}$. Then Conditions {\rm(a)} and
{\rm(b)} are equivalent to
\begin{enumerate}
\item[\rm(i)] $l_2|_{\wedge^2 A} =[\cdot,\cdot]_A$,\quad $l_i|_{\wedge^i A}=0,~i>2$,\quad $\rho|_{A}=\rho_A$, and $\rho|_{E_0}=0$;
\item[\rm(ii)] $l_i(v^k, v^l, \dots)=0$, for $v^k\in \Gamma(E_{-k})$ and $v^l\in
  \Gamma(E_{-l})$;
\item[\rm(iii)]  $l_i(X_1,\dots,X_{i-1},v^k)\in
\Gamma(E_{-k-i+2})$, for $X_j\in \Gamma(A),~j=1,\cdots,i-1,~v^k\in \Gamma(E_{-k})$.
\end{enumerate} The advantage of these criteria is that they apply
to the infinite dimensional case since no dual space is
involved. Moreover, we notice that  the only condition to check in
{\rm (iii)} is
\begin{enumerate}
\item[\rm(iii')] $l_2(X, v^0)\in \Gamma(E_0)$, for $X\in \Gamma(A)$ and $v^0 \in \Gamma(E_0)$;
\end{enumerate}
all the rest automatically hold for degree reasons.
\end{rmk}

\emptycomment{
Now we make the 2-term case more explicit. Given a 2-term representation up to homotopy $(\huaE=E_{-1}\oplus E_0,D)$
of $A$, where $D=\partial+d_\nabla+\omega$ (see Example \ref{pro:rep-homotopy} for details),
 the corresponding Chevalley-Eilenberg complex
$C(A\ltimes \huaE)$
 is
\begin{eqnarray*}
 &&\CWM\longrightarrow \Gamma((A\oplus E_0)^*)\longrightarrow \Gamma( \wedge^2(A\oplus E_0)^*)\oplus
\Gamma(E_{-1}^*)\\
&&\longrightarrow\Gamma( \wedge^3(A\oplus E_0)^*)\oplus
\Gamma((A\oplus E_0)^*)\otimes\Gamma(E_{-1}^*)\longrightarrow\cdots,
\end{eqnarray*}
where $E_0^*$ is of degree 1 and $E_{-1}^*$ is of
degree 2.}

Now we make the brackets and the anchor of the semidirect product Lie 2-algebroid more explicit. Let $\huaE=E_{-1}\oplus E_0$ be a $2$-term graded vector bundle and $(\huaE,D)$ a representation up to homotopy of a Lie algebroid $A$. From Example \ref{pro:rep-homotopy}, we know that $D=\partial +d_\nabla+\omega$.

\begin{pro}\label{pro:li}
Let $(\huaE,D)$ be a $2$-term representation up to homotopy of a Lie algebroid
  $A$, where $D=\partial +d_\nabla+\omega$. Then, the semidirect product Lie $2$-algebroid structure on the graded
  vector bundle $[A\oplus E_0] \oplus E_{-1}$ is given by
 \begin{eqnarray}
\rho(X+u)&=&\rho_A(X), \\
   \label{semidirectproduct1} l_1(m)&=&\partial m,\\
    \label{semidirectproduct2}l_2(X+u,Y+v)&=&[X,Y]_A+\nabla_Xv-\nabla_Yu,\\
    \label{semidirectproduct3}l_2(X+u,m)&=&\nabla_Xm,\\
\label{semidirectproduct4}l_3(X+u,Y+v,Z+w)&=&-\omega(X,Y)(w)-\omega(Y,Z)(u)-\omega(Z,X)(v),
  \end{eqnarray}
 for all $X,Y,Z\in \Gamma (A),~u,v,w\in \Gamma (E_0),~m\in \Gamma (E_{-1}).$
\end{pro}

\pf Let $D^*:C(\huaM)\otimes \Gamma \big(\huaE^* \big)
  \longrightarrow C(\huaM)\otimes \Gamma \big( \huaE^* \big)$ be the
  dual $2$-term representation up to homotopy of $A$ on $\huaE^*$, and $\widetilde{D}:C(\huaM)\otimes \Gamma \big(\Sym(\huaE[1])^* \big)
  \longrightarrow C(\huaM)\otimes \Gamma \big(\Sym( \huaE[1])^* \big)$
  the $2$-term representation up to homotopy of $A$ on $\Sym(
  \huaE[1])^*$. We decompose $D^*=\partial^*+d_{\nabla^*}+\omega^*$, where
  $\nabla^*$ is the dual $A$-connection on $\huaE^*$, $\partial^*$ and $\omega ^*$ are given by
  $\partial^*(\xi^0)(m)=-\xi^0(\partial m)$ and $\omega^*(X,Y)(\xi^1)(u)=\xi^1(\omega(X,Y)(u)),$ for all $\xi^1\in\Gamma(E_{-1}^*), \xi^0\in\Gamma(E_0^*), u\in\Gamma(E_0), m\in\Gamma(E_{-1})$ and $X,Y\in\Gamma(A)$.

 Now we apply the equations in \eqref{eq:q-brackets}
repetitively. For all $f\in \CWM$, by Proposition \ref{thm:Qreputh}, we have
$
\widetilde{D}(f)=\rho^*(df)=d_Af,
$
which implies that
$
\rho(X+u)=\rho_A(X).
$

Similarly, for all $\xi^0\in\Gamma(E_0^*)$, $m\in E_{-1}$,  we have
$
\widetilde{D}(\xi^0)(m)=
D^*(\xi^0)(m)= \partial^*(\xi^0)(m)= \xi^0(-\partial
m).
$
On the other hand, we have
$
\widetilde{D}(\xi^0)(m)= \xi_0(-l_1(m)),
$
which implies that $l_1=\partial$.

For all $X\in\Gamma(A),~u\in\Gamma(E_0)$, we have
$$\widetilde{D}(\xi^0)(X,u)=D^*(\xi^0)(X,u)=d_{\nabla^*}\xi^0(X)(u)=(\nabla^*_X\xi^0)(u)=\rho(X)(\xi^0(u))-\xi^0(\nabla_Xu).$$
On the other hand, we have
$
\widetilde{D}(\xi^0)(X,u)=\rho(X)(\xi^0(u))-\xi^0(l_2(X,u)),
$
which implies that  $l_2(X,u)=\nabla_Xu.$

For all $\phi\in\Gamma(A^*)$, by Proposition \ref{thm:Qreputh}, we have
$ \widetilde{D}(\phi)(X,Y)=d_{A}\phi(X,Y)=\rho(X)\phi(Y)-\rho(Y)\phi(X)-\phi([X,Y]_A)$. On the other hand, we have
$\widetilde{D}\phi(X,Y)=\rho(X)\phi(Y)-\rho(Y)\phi(X)-\phi(l_2(X,Y))$, which implies that  $l_2(X,Y)=[X,Y]_A$.

Similarly, for all $\xi^1\in\Gamma(E_{-1}^*)$, we have
$$
\widetilde{D}(\xi^1)(X,m)=D^*(\xi^1)(X,m)= d_{\nabla^*}\xi^1(X)(m)= \nabla^*_X\xi^1(m)=\rho(X)\xi_1(m)- \xi_1(\nabla_Xm).
$$
Comparing with $
\widetilde{D}(\xi^1)(X,m)=\rho(X)\xi^1(m)- \xi_1(l_2(X,m))
$,
we have $ l_2(X,m)=\nabla_Xm$.

Finally, by comparing
$$
\widetilde{D}(\xi^1)(X,Y, u)=\omega^*(\xi^1)(X, Y,
u)= \omega^*(X,Y)(\xi^1)(u)=\xi^1(\omega(X,Y)(u))
$$
 with
$
\widetilde{D}(\xi^1)(X,Y, u)=(\xi^1,-l_3(X, Y, u)),
$ we see that
$
  l_3(X, Y, u)=-\omega(X,Y)(u)
$,
which implies that
\eqref{semidirectproduct4} holds. The proof is finished.
\qed\vspace{3mm}

Recall from \cite[Example 3.28]{abad-crainic:rep-homotopy} that a
Lie algebroid $A$ has a natural coadjoint representation up to
homotopy on $T^*M \xrightarrow{\rho^*} A^*$, which is given by a choice of
$A$-connection on $T^*M$. Moreover, any  two connections induce
equivalent representations and equivalent representations give rise
to isomorphic semidirect products.    When $A=TM$, we obtain a
representation up to homotopy of $TM$ on $T^*M \xrightarrow{id}
T^*M$, that is, a $TM$-connection $\nabla$ on $T^*M$, and a 2-form
$\omega\in \Omega^2(TM, \End(T^*M, T^*M))$ satisfying
\[ \nabla_{[X, Y]}-[\nabla_X,\nabla_Y]= \omega (X, Y), \quad \forall X, Y \in \Gamma(TM).
\]
Then by Proposition \ref{pro:li}, the Lie 2-algebroid structure on $TM
\ltimes (T^*M \xrightarrow{id} T^*M)$ is given by
 \begin{eqnarray}
\label{courant0} \rho(X+\xi)&=&\rho_A(X) \\
   \label{courant1} l_1(\alpha)&=& \alpha,\\
    \label{courant2}l_2(X+\xi,Y+\eta)&=&[X,Y]+\nabla_X(\eta)-\nabla_Y(\xi),\\
    \label{courant3}l_2(X+\xi,\alpha)&=&\nabla_X(\alpha),\\
\label{courant4}l_3(X+\xi,Y+\eta,Z+\gamma)&=&-\omega(X,Y)(\gamma)-\omega(Y,Z)(\xi)-\omega(Z,X)(\eta),
  \end{eqnarray}
 for all $X,Y,Z\in \Gamma (TM),~\xi,\eta,\gamma,\alpha\in \Gamma (T^*M).$
 Since this is the case relevant to the construction of the Courant
algebroid, we now explicitly write down the isomorphism between the
semidirect products given by two representations up to homotopy
$(\nabla,\omega)$ and $(\nabla^\prime,\omega^\prime)$ of $TM$.  We
assume that $\nabla_X-\nabla^\prime_X=B(X)$ for some bundle map $B:TM\longrightarrow
\Hom(T^*M,T^*M)$.  Then it is not hard
to check that
$f=(f_0,f_1,f_2): TM \ltimes_{\nabla, \omega}
(T^*M\xrightarrow{id}T^*M)\longrightarrow TM \ltimes_{\nabla',
  \omega'} (T^*M\xrightarrow{id}T^*M) $ defined by
\begin{eqnarray*}
f_0&=&{id}:TM\oplus T^*M\longrightarrow TM\oplus T^*M,\\
f_1&=&{id}:T^*M\longrightarrow T^*M,\\
f_2(X+\xi,Y+\eta)&=&B(X)(\eta)-B(Y)(\xi).
\end{eqnarray*}
preserves the anchor and the brackets in an $L_\infty$-fashion (see
Remark \ref{remark:isom-lie-n}).  Thus, it is  an isomorphism of Lie
2-algebroids. Therefore, if we only care about the isomorphism
class, we can take {\em the} semidirect product $TM \ltimes
(T^*M\xrightarrow{id} T^*M)$.

\emptycomment{
The symplectic NQ-manifold associated to the standard Courant
algebroid $TM\oplus T^*M$ is $T^*[2]T[1]M$. The symplectic structure
is the standard one: $dp_i\wedge dq^i+d\xi^i\wedge d\theta_i$ and
the degree 1 vector field $Q$ is given by
}
In the following, we show that the standard Courant
algebroid is a semidirect product with the help of Proposition
\ref{thm:Qreputh}.

\begin{thm}\label{thm:Qsatisfycondition}
The
semidirect product $TM\ltimes (T^*M\xrightarrow{id} T^*M)$ gives rise
to the underlying degree $2$ NQ manifold $T^*[2]T[1]M$ of the standard
Courant algebroid.
\end{thm}
\pf Recall from Section \ref{sec:courant}, in local coordinates therein, the
homological vector field of the standard Courant algebroid
$T^*[2]T[1]M$ is
\begin{equation}\label{eq:standardQ}
Q=\xi^i\frac{\partial}{\partial q^i}+p_i\frac{\partial}{\partial
\theta_i}.
\end{equation}
 Thus we only need to show that this $Q$ satisfies Conditions
{\rm (a)-(b)} listed in Proposition \ref{thm:Qreputh}. For all $f\in
\CWM$, we have $ Q(f)=\xi^i\frac{\partial f}{\partial q^i}=df. $ For
all $\xi=f_j\xi^j\in\Gamma(T^*[1]M)$, with $f_j\in \CWM$, we have
$$
Q(\xi)=\xi^i\frac{\partial\xi}{\partial q^i}=\frac{\partial
f_j}{\partial q^i}\xi^i\xi^j=d\xi \in \Omega^2 (M).
$$
Then the fact that $Q$ satisfies Condition (a) in Proposition
\ref{thm:Qreputh} follows from the derivation property of  $Q$.

For all $\theta = f^j \theta_j \in \Gamma(T[1]M)$, with $f^j\in
\CWM$, we have
$$
Q(\theta)=\xi^i\frac{\partial \theta}{\partial
q^i}+p_i\frac{\partial\theta}{\partial\theta_i}=\xi^i\frac{\partial
f^j}{\partial
q^i}\theta_j+f^jp_i\frac{\partial\theta_j}{\partial\theta_i}=\frac{\partial
f^j}{\partial q^i}\xi^i\theta_j+f^ip_i.
$$
For all $p=f^j p_j\in\Gamma(T^*[2]M)$, we have $
Q(p)=\xi^i\frac{\partial p}{\partial q^i}=\frac{\partial
f^j}{\partial q^i}\xi^ip_j. $ Thus Condition (b) in Proposition
\ref{thm:Qreputh} is also satisfied. \qed

\begin{rmk}\label{rmk:D}
 In local coordinates $(q^i,\xi^i,p_i,\theta_i)$, we are free to choose the flat
$TM$-connection $\nabla$ on $T^*M$ given by $ \nabla_{\theta_i}\xi^j=0.$
Since it is flat,   $\omega=0$. It is straightforward to see that
in this case, $D=\partial+d_\nabla+\omega=id +$$d_\nabla$ is exactly
given by $\xi^i\frac{\partial}{\partial
q^i}+p_i\frac{\partial}{\partial \theta_i}$.
\end{rmk}

\begin{rmk} For $r> 2$,  $T^*[r] T[1]M$
(see  \cite[Section 8]{zambon:l-infty})  can be similarly realized as a Lie $r$-algebroid of the semidirect
  product of $TM$ with $T^*M \xrightarrow{id} T^*M \to 0 \dots\to 0$
  where $T^*M$ lives in degrees $-r+1$ and $-r+2$. With a shifted
  version of the coordinate functions in Section \ref{sec:courant},  the canonical symplectic form $dp_i  dq^i+d\xi^i
d\theta_i$ is of degree $r$, the Hamiltonian function $p_i\xi^i$ is of degree $r+1$ and $Q=\xi^i\frac{\partial}{\partial q^i}+(-1)^rp_i\frac{\partial}{\partial
\theta_i}.$
\end{rmk}

Using Getzler's higher derived brackets \cite{getzler:higher-derived}, Zambon constructed  a Lie $r$-algebra
in \cite[Section 8]{zambon:l-infty}.  This Lie $r$-algebra can be viewed as a higher extension of the
Lie 2-algebra related to a Courant algebroid constructed in
\cite{rw}. Now we realize this Lie $r$-algebra  as a semidirect
product. Thus, our observation here is a higher
analogue of the example we gave in \cite{shengzhu1}.
\begin{ep}{\rm
Given a manifold $M$ and an integer $r\ge 2$, we recall from
\cite[Proposition 8.1]{zambon:l-infty} the Lie $r$-algebra structure on
\[ C^\infty(M) \xrightarrow{d} \dots \xrightarrow{d} \Omega^{r-2}(M)
\xrightarrow{d} \frkX(M) \oplus \Omega^{r-1}(M),   \]
\begin{enumerate}
\item[$\bullet$] $l_1$ is given by de Rham differential $d$;
\item[$\bullet$] $l_2$ is given by
 \begin{eqnarray*}
 l_2((X, \xi), (Y,
  \eta))&=&[X,Y]+L_{X}\eta-L_{Y}\xi-\half
  d(i_{X}\eta-i_{Y}\xi),\\
  l_2((X, \xi),
  \alpha)&=&\half L_X \alpha,
\end{eqnarray*}
 for all $(X, \xi)$, $(Y,
  \eta)\in \frkX(M) \oplus \Omega^{r-1}(M)$ and  $\alpha \in \Omega^{i(<r-1)}(M)$;
\item[$\bullet$] $l_3$ is given by
\begin{eqnarray*}
  l_3(e_1,e_2,e_3)&=&-\frac{1}{6}\big(\pair{l_2(e_1,e_2),e_3}+c.p.\big),\\
  l_3(\alpha,e_1,e_2)&=&-\frac{1}{6}\Big(\frac{1}{2}(i_{X_1}L_{X_2}-i_{X_2}L_{X_1})+i_{[X_1,X_2]}
  \Big)\alpha,
\end{eqnarray*}
for all $e_1=(X_1,\xi_1),e_2=(X_2,\xi_2)\in \frkX(M) \oplus
\Omega^{r-1}(M)$ and $\alpha\in\Omega^{i(<r-1)}(M)$, where the
pairing $\pair{\cdot,\cdot}:(TM\oplus \wedge^{r-1}T^*M)\times
(TM\oplus \wedge^{r-1}T^*M)\longrightarrow \wedge^{r-2}T^*M$ is
given by
$$
\pair{(X,\xi),(Y,\eta)}=i_{X}\eta+i_Y\xi.
$$
\item[$\bullet$] $l_n$,  for $n>3$ with $n$ an odd integer, is given by
\begin{eqnarray*}
  l_n(e_1,\cdots,e_n)&=&\sum_il_n(X_1,\cdots,\xi_i,\cdots,X_n)
\end{eqnarray*}
with
$$
l_n(\xi,X_1,\cdots,X_{n-1})=\frac{(-1)^{\frac{n+1}{2}}12B_{n-1}}{(n-1)(n-2)}\sum_{i<j}(-1)^{i+j+1}i_{X_{n-1}}\cdots
\widehat{i_{X_j}}\cdots\widehat{i_{X_i}}\cdots i_{X_1}l_3(\xi,X_i,X_j),
$$
and
$$
l_n(\alpha,X_1,\cdots,X_{n-1})=\frac{(-1)^{\frac{n+1}{2}}12B_{n-1}}{(n-1)(n-2)}\sum_{i<j}(-1)^{i+j+1}i_{X_{n-1}}\cdots
\widehat{i_{X_j}}\cdots\widehat{i_{X_i}}\cdots i_{X_1}l_3(\alpha,X_i,X_j),
$$
where $e_i=(X_i,\xi_i)\in\frkX(M)\oplus\Omega^{r-1}(M)$,
$\xi\in\Omega^{r-1}(M)$ and $\alpha\in\Omega^{i(<r-1)}(M)$. Here the
$B$'s denote the Bernoulli numbers.

\end{enumerate}
Take $A=\frkX(M)$ and $\huaE_r= C^\infty(M) \oplus \Omega^1(M)
\oplus \dots \oplus \Omega^{r-1}(M)$. It is easy to see that the Lie
$r$-algebra structure satisfies the three conditions in Remark
\ref{rmk:dual-semidirect-product}. Thus, this Lie $r$-algebra is
simply $\frkX(M) \ltimes \big( C^\infty(M) \xrightarrow{d}
\Omega^1(M) \xrightarrow{d} \dots \xrightarrow{d} \Omega^{r-1}(M)
\big) $. The complexity of the brackets gives us a very interesting
$L_\infty$-module of $\frkX(M)$.}
\end{ep}

\section{Extensions of Lie 2-algebroids} \label{sec:ext}

Recall that there is a one-to-one correspondence between  Lie algebra
extensions of a Lie algebra $\frkg$ by a $\g$-module $V$ and $H^2(\frkg, V)$ (which includes the information of
$V$ as a $\frkg$-module). Given a split Lie $n$-algebroid $(\huaA,
\{l_i\}_{i=1}^{n+1}, \rho)$  and a representation up to homotopy $(\huaE, D)$, we have
a very natural cohomology group  $H^\bullet(\huaA, \huaE)$
associated to this representation up to homotopy, by taking the cohomology of the complex $D:
C(\huaA)\otimes \Gamma(\huaE) \to C(\huaA)\otimes \Gamma(\huaE)$.
Thus we might ask
the following questions:
\begin{itemize}
 \item What kind of extensions up to isomorphisms may be measured by  the
cohomology group $H^\bullet(\huaA, \huaE)$?

\item Will these   extensions contain nice
examples such as Courant algebroids?
\end{itemize}
The answer of the first question is given exactly by the abelian extensions that we will define in the sequel.   Furthermore,  it
turns out that all the examples we care in this paper are covered by this kind of extensions.

Now we consider   extensions of Lie 2-algebroids of the
following form:
\begin{equation}\label{eq:ext-a}
\CD
  0 @>0>>  E_{-1} @>id_{-1}>> \hA_{-1}=E_{-1} @>0>> 0 @>0>> 0 \\
  @V 0 VV @V \partial VV @V \widehat{\partial} VV @V0 VV @V0VV  \\
  0 @>0>> E_{0} @>\inc>> \hA_0 @>\p>> A@>0>>0
\endCD
\end{equation} The exact sequence consists of two exact sequences of
vector bundles with all squares commutative diagrams of vector
bundles over the same base $M$. Moreover, $A$ is a Lie algebroid viewed as the Lie 2-algebroid
$0\stackrel{0}{\longrightarrow}A$ as in Example
 \ref{ep:lower-algd},
$E_{-1}\stackrel{\partial}{\longrightarrow}E_0$ is a 2-term complex
of vector bundles viewed as an abelian Lie
 2-algebroid as in Example \ref{ep:ab-algd}, $(\hA_{-1}
 \xrightarrow{\widehat{\partial}} \hA_0, \hrho, \hl_2, \hl_3)$ is a Lie 2-algebroid, and  $(id_{-1},\inc)$ and $(0,\p)$ are strict
 morphisms (see Remark \ref{remark:isom-lie-n})  over $id_M$ between Lie
 2-algebroids.  We call this sort of extension a {\bf $2$-term abelian
 extension of a Lie algebroid $A$} if
 \begin{equation}\label{eq:defiext}
 \hl_3( \inc(u), \inc(v),\cdot)=0,  \quad \forall   u,
v \in \Gamma(E_0).
 \end{equation}

 We will simply write
$\huaE\longrightarrow \widehat{A}\longrightarrow A$ as an abelian
extension of a Lie algebroid $A$ by a 2-term complex
$\huaE:E_{-1}\stackrel{\partial}{\longrightarrow}E_0$.

 Let $\huaE\longrightarrow \widehat{A}\longrightarrow A$ and $\huaE\longrightarrow \widetilde{ A}\longrightarrow A$ be two abelian extensions of
 $A$ by the 2-term complex $\huaE:E_{-1}\stackrel{\partial}{\longrightarrow}E_0$. An {\bf isomorphism} between them is a Lie
 2-algebroid isomorphism $f=(f_0,f_1,f_2):\widehat{A}\longrightarrow\widetilde{ A}$  (not necessarily strict, see Definition \ref{def:morphism} and
 Remark \ref{remark:isom-lie-n}) satisfying
\begin{equation}\label{eq:defiiso}
f_2(\inc(u),\cdot)=0, \quad \forall\; u\in \Gamma(E_0),
\end{equation}
 and the following commutative diagram:
 \[
\xymatrix{ \huaE\ar@{=}[d] \ar[r] & \widehat{A}  \ar@{->}^{f}[d]  \ar[r] & A\ar@{=}[d] \\
\huaE  \ar[r] & \widetilde{ A}\ar[r] & A.   }
\]

\emptycomment{However, it turns out that already in the case of $L_\infty$-algebras, one needs to
go beyond this sort of cohomology (see
\cite[Section 3]{lazarev:extension}, also see Remark
\ref{rmk:cohomology}). Nevertheless, we now show that our naive generalization
$H^\bullet(\huaA, \huaE)$ still gives arise to a partial one-to-one
correspondence which is enough for our application in this paper. The
advantage of this cohomology is that the cochain complex is relatively
small thus easy to calculate with explicit formulas.
}

\emptycomment{
Moreover we notice that $\p$ admits a splitting
 $\sigma: A \to \hA_0$ by  partitions of unity, so we can write
 $\hA_{0}=A\oplus E_0$. The fact that $(id_{-1},\inc)$ and $(0,\p)$
 preserve the brackets implies that
\[\hl_2(\hl_2(\sigma(X), u), v) +\hl_2(\hl_2(u, v), \sigma(X)) +\hl_2(\hl_2(v, \sigma(X)), u)   =0, \quad \forall X \in \Gamma(A), u,
v \in \Gamma(E_0).
\]
Therefore, $\widehat{\partial}\hl_3(\sigma(X), u, v)=0$.
Notice that if we have another splitting $\sigma': A\to \hA_0$, then
the difference $\sigma(X)-\sigma'(X)$ is  in $E_0$. Thus a different choice
of splitting does not affect the definition of an abelian extension because $\hl_3$
restricting on $\wedge^3E_0$ is 0.
}

Given an extension \eqref{eq:ext-a}, we notice that $\p$ admits a splitting
 $\sigma: A \to \hA_0$ by  partitions of unity, so we can write
 $\hA_{0}=A\oplus E_0$.  Since
the diagram is commutative, we have that
$\hat{\partial}=0+\partial,~\hrho=\rho_A$ and $\inc$ is the
inclusion map, which we usually omit. Define an $A$-connection
$\nabla=(\nabla^{-1},\nabla^0)$ on the complex $E_{-1}\stackrel{\partial}{\longrightarrow}E_0$ and $\omega:\wedge^2A
\longrightarrow\End(E_0,E_{-1})$ by
\begin{eqnarray*}
 \nabla^0_X(u)&=&\hl_2(\sigma(X),u),\\
 \nabla^{-1}_X(m)&=&\hl_2(\sigma(X),m),\\
\omega(X,Y)(u)&=&-\hl_3(\sigma(X),\sigma(Y),u),
\end{eqnarray*}
and define  $c_2\in\Gamma(\Hom(\wedge^2 A,E_0))$ and
$c_3\in \Gamma(\Hom (\wedge^3A, E_{-1}))$ by
\[
c_2(X,Y)=\hl_2(\sigma(X),\sigma(Y))-\sigma([X,Y]_A), \quad
c_3(X,Y,Z)=\hl_3 (\sigma(X),\sigma(Y),\sigma(Z)),
\]
for all $X,Y,Z \in \Gamma(A),~u\in \Gamma(E_0), ~m\in \Gamma(E_{-1}).$

\begin{lem} \label{lemma:ext-rep}
 With the above notations, $(\nabla,\omega)$ gives a representation up to homotopy of the Lie
 algebroid
$A$ on the complex
$\huaE:E_{-1}\stackrel{\partial}{\longrightarrow}E_0.$
\end{lem}
An extension of $A$ is called {\bf
 central}  if $(\nabla, \omega)=0$.

\pf It is not hard to deduce that
\begin{eqnarray*}
 && [\nabla_X,\nabla_Y](u)-\nabla_{[X,Y]_A}u\\&=&\hl_2(\sigma(X),\hl_2(\sigma(Y),u))-\hl_2(\sigma(Y),\hl_2(\sigma(X),u))-\hl_2(\sigma([X,Y]_A),u)\\
  &=&\hl_2(\sigma(X),\hl_2(\sigma(Y),u))-\hl_2(\sigma(Y),\hl_2(\sigma(X),u))\\&&-\hl_2(\hl_2(\sigma(X),\sigma(Y)),u)
  +\hl_2(c_2(X,Y),u).
\end{eqnarray*}
Thus, we have
$$
[\nabla_X,\nabla_Y](u)-\nabla_{[X,Y]_A}u=\hat{\partial}
\hl_3(\sigma(X),\sigma(Y),u)=-\partial( \omega(X,Y)(u)).
$$
Similarly, we have
$$
[\nabla_X,\nabla_Y](m)-\nabla_{[X,Y]_A}m=
\hl_3(\sigma(X),\sigma(Y),\partial m)=- \omega(X,Y)(\partial m).
$$
By \eqref{eq:defiext}, we have
\begin{eqnarray*}
[\nabla_X,\omega(Y,Z)](u)&=&\nabla_X\omega(Y,Z)(u)-\omega(Y,Z)(\nabla_Xu)\\
&=&-\hl_2(\sigma(X),\hl_3(\sigma(Y),\sigma(Z),u))+\hl_3(\sigma(Y),\sigma(Z),\hl_2(\sigma(X),u)),\\
\omega([X,Y]_A,Z)(u)&=&-\hl_3(\sigma([X,Y]_A),\sigma(Z),u)\\
&=&-\hl_3(\hl_2(\sigma(X),\sigma(Y))-c_2(X,Y),\sigma(Z),u)\\
&=&-\hl_3(\hl_2(\sigma(X),\sigma(Y)),\sigma(Z),u).
\end{eqnarray*}
Since
$
\hl_2(\hl_3(\sigma(X),\sigma(Y),\sigma(Z)),u)=0,
$
by the Jacobiator identity involving $\hl_3$, we deduce that
\begin{eqnarray*}
&& [\nabla_X,\omega(Y,Z)]-
\omega([X,Y]_A,Z)+c.p.(X,Y,Z)=0,
\end{eqnarray*}
which implies that $(\nabla,\omega)$ is a representation up to
homotopy of $A$ on $E_{-1}\stackrel{\partial}{\longrightarrow}E_0.$
\qed \vspace{3mm}

Now we recall how to define the cohomology $H^\bullet (A, \huaE)$ for
a 2-term representation up to homotopy $(\huaE=E_{-1} \oplus E_0, D)$ of a Lie algebroid
$A$. Such a representation up to homotopy gives us a
complex:
\begin{equation} \label{eq:cx-coh-algd}
\begin{split}
&\Gamma(E_{-1})\stackrel{D}{\longrightarrow}\Gamma(E_0\oplus
 \Hom(A,E_{-1}))\stackrel{D}{\longrightarrow}\Gamma( \Hom(A,E_0)\oplus \Hom(\wedge^2A,E_{-1}))
 \stackrel{D}{\longrightarrow} \\ &\Gamma(\Hom(\wedge^2A,E_0)\oplus \Hom(\wedge^3A,E_{-1}))
 \stackrel{D}{\longrightarrow} \Gamma(\Hom(\wedge^3A,E_0)\oplus
 \Hom(\wedge^4A,E_{-1}) )\longrightarrow\cdots.
\end{split}
\end{equation}
 We
write $D=\partial+d_\nabla+\omega$ according to Example
\ref{pro:rep-homotopy}. Then  for any $k$-cochain
$$(\varpi_1,\varpi_2)\in\Gamma(\Hom(\wedge^k
A,E_0)\oplus\Hom(\wedge^{k+1}A,E_{-1})),$$ we have
\begin{equation}\label{eqn:D}
D(\varpi_1,\varpi_2)=(d_\nabla\varpi_1+ \partial(\varpi_2),d_\nabla\varpi_2+\omega
(\varpi_1)).
\end{equation}
This complex eventually gives us the cohomology $H^\bullet(A,
\huaE)$ with the coefficient in the representation up to homotopy
$\huaE$.

In particular, for a 2-cochain
$(c_2,c_3)\in\Gamma(\Hom(\wedge^2A,E_0)\oplus\Hom(\wedge^3A,E_{-1}))$, we
have
$$
D(c_2,c_3)=(d_\nabla c_2+\partial ( c_3))+(d_\nabla
c_3+\omega(
c_2))\in\Gamma(\Hom(\wedge^3A,E_0)\oplus\Hom(\wedge^4A,E_{-1})).
$$
Thus, $(c_2,c_3)$ is a 2-cocycle if and only if
$$d_\nabla c_2+\partial ( c_3)=0,\quad d_\nabla c_3+\omega(c_2)=0.$$

Transferring the Lie 2-algebroid structure $(\hA_{-1}
 \xrightarrow{\hl_1=\widehat{\partial}} \hA_0, \hrho, \hl_2, \hl_3)$
 to the isomorphic   complex $E_{-1}\xrightarrow{0+\partial}A\oplus E_0$, we obtain the Lie 2-algebroid
 $(E_{-1}\xrightarrow{0+\partial}A\oplus E_0,\rho,l_2,l_3)$, where
\begin{equation} \label{eq:ext-a-structure}
\left\{\begin{array}{rcl}
\rho(X+u)&=&\rho_A(X),\\
l_2(X+u,Y+v)&=&[X,Y]_A+\nabla_Xv-\nabla_Yu+c_2(X,Y),\\
l_2(X+u,m)&=&\nabla_Xm,\\
l_3(X+u,Y+v,Z+w)&=&c_3(X,Y,Z)-\big(\omega(X,Y)(w)+\omega(Y,Z)(u)+\omega(Z,X)(v)\big).
\end{array}\right.
\end{equation}

\begin{lem}\label{lemma:ext-cocycle}
With the above notations, the fact that Eq. \eqref{eq:ext-a-structure}
gives a Lie $2$-algebroid structure implies that  $(c_2,c_3)$ is a $2$-cocycle of  the Lie
algebroid $A$ with coefficients in the representation up to homotopy
$(E_{-1}\stackrel{\partial}{\longrightarrow}E_0,\nabla,\omega)$ (see
the complex \eqref{eq:cx-coh-algd}).

Conversely, given a $2$-term
representation up to homotopy
$(E_{-1}\stackrel{\partial}{\longrightarrow}E_0,\nabla,\omega)$   of
$A$ and a $2$-cocycle $(c_2,c_3)\in \Hom(\wedge^2A,E_0)\oplus
\Hom(\wedge^3A,E_{-1})$,  defining $\rho,l_2,l_3$ by
\eqref{eq:ext-a-structure},  we obtain a Lie $2$-algebroid,
$$
A\ltimes_{(c_2,c_3)}\huaE:=(E_{-1}\xrightarrow{0+\partial}A\oplus
E_0,\rho,l_2,l_3),
$$
 which is a $2$-term abelian extension of $A$
fitting into
 \eqref{eq:ext-a}.
\end{lem}
\pf  Assume that Eq. \eqref{eq:ext-a-structure} gives rise to a Lie
2-algebroid structure. Then we have
\begin{eqnarray*}
&&l_2(Z+w,l_2(X+u,Y+v))+c.p.(Z+w, X+u,Y+v)\\&=&\partial
l_3(Z+w,X+u,Y+v)\\
&=&\partial
c_3(X,Y,Z)-\partial\big(\omega(X,Y)(w)+\omega(Y,Z)(u)+\omega(Z,X)(v)\big).
\end{eqnarray*}
On the other hand, by direct computations, we have
\begin{eqnarray*}
&&l_2(Z+w,l_2(X+u,Y+v))+c.p.(Z+w, X+u,Y+v)\\&=&l_2(Z+w,[X,Y]_A+\nabla_Xv-\nabla_Yu+c_2(X,Y))+c.p.(Z+w, X+u,Y+v)\\
&=&-\nabla_{[X,Y]_A}w+\nabla_Z\nabla_Xv-\nabla_Z\nabla_Yu+\nabla_Zc_2(X,Y)\\&&+c_2(Z,[X,Y]_A)+c.p.(Z+w, X+u,Y+v)\\
&=&d_\nabla
c_2(X,Y,Z)-\partial\big(\omega(X,Y)(w)+\omega(Y,Z)(u)+\omega(Z,X)(v)\big).
\end{eqnarray*}
Thus, we have $d_\nabla
c_2(X,Y,Z)-\partial
c_3(X,Y,Z)=0$, i.e.
\begin{equation}\label{omegatheta1}
d_\nabla c_2+\partial (c_3)=0.
\end{equation}

 By \eqref{eq:l-infty-brackets}, we have
\begin{eqnarray*}
  &&l_2(X+u,l_3(Y+v,Z+w,P+t))-l_2(Y+v,l_3(X+u,Z+w,P+t))\\
  &&+l_2(Z+w,l_3(X+u,Y+v,P+t))-l_2(P+t ,l_3(X+u,Y+v,Z+w))\\
  & =&l_3(l_2(X+u,Y+v),Z+w,P+t)-l_3(l_2(X+u,Z+w), Y+v,P+t)\\
  &&+l_3(l_2(X+u,P+t ),Y+v,Z+w)+l_3(l_2(Y+v, Z+w),X+u, P+t)\\
  &&-l_3(l_2(Y+v, P+t),X+u,Z+w )+l_3(l_2(Z+w, P+t),X+u,Y+v).
\end{eqnarray*}
\emptycomment{
which implies that
$$
\nabla_Xc_3(Y,Z,P)-\nabla_X\big(\omega(Y,Z)(t)+\omega(Z,P)(v)+\omega(P,Y)(w)\big)+c.p.
$$
\comment{c.p. of what? $X, Y, Z, P$? then probably it's unshuffle. or
  it also involves $t, v, w$? it's not very clear to me... also why
  only $t, v, w$ but no $u$? it's all not very clear.}
is equal to
\begin{eqnarray*}
c_3([X,Y]_A,Z,P)-\omega(Z,P)\big(c_2(X,Y)+\nabla_Xv-\nabla_Yu\big)-\omega([X,Y]_A,Z)(t)-\omega(P,[X,Y]_A)(w)+c.p..
\end{eqnarray*}
\comment{the same thing, please be more clear about c.p.}
}
By \eqref{eq:ext-a-structure} and the fact that $(\nabla,\omega)$ is a representation up to
homotopy, we deduce that
\begin{equation}\label{omegatheta2}
d_\nabla c_3+\omega( c_2)=0.
\end{equation}
By \eqref{omegatheta1} and \eqref{omegatheta2}, we deduce that
$(c_2,c_3)$ is a 2-cocycle. It is straightforward to check that the
converse part also holds. The proof is completed.
 \qed\vspace{3mm}

Denote by $\widetilde{c_2}:\Gamma(E_0^*)\longrightarrow
\Gamma(\wedge^2 A^*)$ and
$\widetilde{c_3}:\Gamma(E_{-1}^*)\longrightarrow \Gamma(\wedge^3
A^*)$ the dual of $c_2$ and $c_3$ respectively, i.e.
\begin{equation}\label{eq:ctilde}
\widetilde{c_2}(\xi^0)(X,Y)=-\xi^0(c_2(X,Y)),\quad
\widetilde{c_3}(\xi^1)(X,Y,Z)=\xi^1(c_3(X,Y,Z)),
\end{equation}
for all $\xi^0\in \Gamma(E_0^*),\xi^1\in \Gamma(E_{-1}^*)$ and
$X,Y,Z\in\Gamma(A)$.
 We denote their graded
derivation extension on $C((A\oplus \huaE)[1])$ by the same notation.  Then, we have

\begin{pro}\label{pro:ext-a-nq}
The Chevalley-Eilenberg complex $C\big( (A\oplus \huaE)[1] \big)$ associated to
 $A\ltimes_{(c_2,c_3)}\huaE$ with the Lie
$2$-algebroid structure \eqref{eq:ext-a-structure} is given by
\begin{equation} \label{eq:d-hat}
\begin{split}
&\CWM\stackrel{\widehat{D}}{\longrightarrow} \Gamma((A\oplus
E_0)^*)\stackrel{\widehat{D}}{\longrightarrow} \Gamma(
\wedge^2(A\oplus E_0)^*)\oplus
\Gamma(E_{-1}^*)\\
&\stackrel{\widehat{D}}{\longrightarrow}\Gamma( \wedge^3(A\oplus
E_0)^*)\oplus \Gamma((A\oplus
E_0)^*)\otimes\Gamma(E_{-1}^*)\stackrel{\widehat{D}}{\longrightarrow}\cdots,
\end{split}
\end{equation}
in which $\Gamma(A)$ and $\Gamma(E_0^*)$ are of degree $1$ and $\Gamma(E_{-1}^*)$ is of
degree $2$, and $\widehat{D}$ is given by
$$
\widehat{D}=\widetilde{D}+\widetilde{c_2}+\widetilde{c_3},
$$
with $\widetilde{D}$ the induced
representation up to homotopy on $\Sym((\huaE[1])^*)$.
\end{pro}

\pf 
For all $\xi^0\in \Gamma(E_0^*)$ and $X,Y\in\Gamma(A)$, we have
$$
\widehat{D}(\xi^0)(X,Y)=\langle\xi^0,-l_2(X,Y)\rangle=\langle\xi^0,-c_2(X,Y)\rangle,
$$
which implies that
$\widehat{D}(\xi^0)(X,Y)=\widetilde{c_2}(\xi^0)(X,Y).$
Similarly, for any $\xi^1\in E_{-1}^*$, we have
$
\widehat{D}(\xi^1)(X,Y,Z)=\langle\xi^1,l_3(X,Y,Z)\rangle=\langle\xi^1,c_3(X,Y,Z)\rangle.
$
Thus we have
$
\widehat{D}(\xi^1)(X,Y,Z)=\widetilde{c_3}(\xi^1)(X,Y,Z).
$
Therefore $
\widehat{D}=\widetilde{D}+\widetilde{c_2}+\widetilde{c_3}.
$
Since $\widehat{D}$ is given by a Lie 2-algebroid structure, it must
satisfy $\widehat{D}^2=0$. \qed\vspace{3mm}

Give a representation up to homotopy $(\huaE,\nabla,\omega)$ and a
2-cocycle $(c_2, c_3)$, by Lemma \ref{lemma:ext-cocycle}, we can
construct an extension of Lie 2-algebroid
$A\ltimes_{(c_2,c_3)}\huaE$. We now prove that the extension
does not depend on the cocycle itself but only on its cohomology class in
$H^2(A, \huaE)$.

\begin{pro} \label{pro:coh}
If two $2$-cocycles $(c_2, c_3)$ and $(c'_2, c'_3)$
represent the same cohomology class in $H^2(A, \huaE)$, then  the
corresponding extensions $A\ltimes_{(c_2,c_3)}\huaE$ and
$A\ltimes_{(c_2',c_3')}\huaE$ are isomorphic.

Conversely, if extensions
$A\ltimes_{(c_2,c_3)}\huaE$ and $A\ltimes_{(c_2',c_3')}\huaE$ are
isomorphic, then $(c_2, c_3)$ and $(c'_2, c'_3)$ represent the same
cohomology class.
\end{pro}

\pf Assume that $2$-cocycles  $(c_2, c_3)$ and $(c'_2, c'_3)$
represent the same cohomology, then we have $(c_2,
c_3)=(c_2',c_3')+D(e_1,e_2)$, for some $(e_1,e_2)\in\Gamma(
\Hom(A,E_0)\oplus\Hom(\wedge^2A,E_{-1}))$. More precisely,
$$
c_2=c_2'+d_\nabla e_1+\partial( e_2),\quad c_3=c_3'+d_\nabla
e_2+\omega( e_1).
$$
Define
$(f_0,f_1):A\ltimes_{(c_2,c_3)}\huaE\longrightarrow A\ltimes_{(c'_2,
c'_3)}\huaE$, and $f_2:\wedge^2(A\oplus E_0)\longrightarrow E_{-1}$ by
\begin{eqnarray*}
  f_0(X+u)=X+u+e_1(X),\quad
  f_1(m)=m, \quad f_2(X+u,Y+v)=e_2(X,Y),
\end{eqnarray*}
for all $X,Y\in\Gamma(A),~u,v\in\Gamma(E_0)$ and $m\in\Gamma(E_{-1})$.
It is clear that $(f_0, f_1)$ is an isomorphism of complexes of
vector bundles and $f_2(u, \cdot)=0$. Thus we only need to show that  $(f_0,f_1,f_2)$
preserve the anchor and the brackets in an $L_\infty$-fashion (see
Remark \ref{remark:isom-lie-n}). First we have
$$
\rho(X+u)=\rho_A(X)=\rho'(X+u+e_1(X))=\rho'(f_0(X+u)),
$$
which implies that $f_0$ preserves the anchor. Moreover, we
have
\begin{eqnarray*}
  &&l_2'(f_0(X+u),f_0(Y+v))\\&=&l_2'(X+u+e_1(X),Y+v+e_1(Y))\\
  &=&[X,Y]_A+\nabla_X(v)-\nabla_Y(u)+\nabla_X(e_1(Y))-\nabla_Y(e_1(X))+c_2'(X,Y),
  \end{eqnarray*}
  and
  \begin{eqnarray*}
  f_0l_2(X+u,Y+v)=[X,Y]_A+\nabla_X(v)-\nabla_Y(u)+e_1([X,Y]_A)+c_2(X,Y).
\end{eqnarray*}
Thus, we have
\begin{eqnarray*}
f_0l_2(X+u,Y+v)-l_2'(f_0(X+u),f_0(Y+v))&=&-d_\nabla e_1(X,Y)+(c_2-c_2')(X,Y)\\
 \nonumber &=&\partial e_2(X,Y)\\
 \nonumber &=&\partial f_2(X+u,Y+v).
\end{eqnarray*}
Similarly, we have
\begin{eqnarray*}
 f_1l_2(X+u,m)- l_2'(f_0(X+u),f_1(m))&=&\nabla_X(m)-\nabla_X(m)=0\\
  &=&f_2(X+u,\partial m).
\end{eqnarray*}
By computations, we have
\begin{eqnarray*}
 && l_2'(f_0(X+u),f_2(Y+v,Z+w))- f_2(l_2(X+u,Y+v),Z+w)+c.p.(X+u,Y+v,Z+w)\\
  &=&l_2'(X+u+e_1(X),e_2(Y,Z))- f_2([X,Y]_A+\nabla_X(v)-\nabla_Y(u)+c_2(X,Y),Z+w)\\
  &&+c.p.(X+u,Y+v,Z+w)\\
  &=&\nabla_Xe_2(Y,Z)- e_2([X,Y]_A,Z)+c.p.(X,Y,Z)\\
  &=&d_\nabla e_2(X,Y,Z).
\end{eqnarray*}
On the other hand, we have
\begin{eqnarray*}
&& f_1 l_3(X+u,Y+v,Z+w)-l_3'(f_0(X+u),f_0(Y+v),f_0(Z+w))\\
&=& c_3(X,Y,Z)-c_3'(X,Y,Z)+\Big(\omega(X,Y)(e_1(Z))+c.p.(X,Y,Z)\Big)\\
&=&c_3(X,Y,Z)-c_3'(X,Y,Z)-\omega(e_1)(X,Y,Z)\\
&=&d_\nabla e_2(X,Y,Z).
\end{eqnarray*}
Thus, $(f_0,f_1,f_2)$ is an isomorphism from the Lie 2-algebroid
$A\ltimes_{(c_2,c_3)}\huaE$ to $A\ltimes_{(c_2',c_3')}\huaE$.
Furthermore, it is obvious that the corresponding extensions are
also isomorphic.

Conversely, given two 2-cocycles $(c_2,c_3)$ and $(c_2',c_3')$, let
$(f_0,f_1,f_2)$ be an isomorphism of the resulting extensions, we
can assume that
$$
f_0(X+u)=X+e_1(X)+u,\quad f_1(m)=m,
$$
for some $e_1\in\Gamma(\Hom(A,E_0))$.
By computations, we have
\begin{eqnarray*}
  l_2'(f_0(X+u),f_0(Y+v))=[X,Y]_A+\nabla_Xe_1(Y)+\nabla_Xv-\nabla_Ye_1(X)-\nabla_Yu+c_2'(X,Y),
\end{eqnarray*}
and
\begin{eqnarray*}
  f_0(l_2(X+u,Y+v))=[X,Y]_A+e_1([X,Y]_A)+\nabla_Xv-\nabla_Yu+c_2(X,Y).
\end{eqnarray*}
By \eqref{eq:l2l'2}, we obtain
\begin{eqnarray}
  \label{eq:iso3}c_2(X,Y)-c_2'(X,Y)&=&(d_\nabla e_1)(X,Y)+\partial
  f_2(X,Y).
\end{eqnarray}
\emptycomment{
\begin{eqnarray}
  \label{eq:iso4}\partial f_2(X,v)&=&0.
\end{eqnarray}
Similarly, by \eqref{eq:f2l2}, we get
\begin{equation}
  \label{eq:iso6}f_2(X,\partial m)=0, \quad\forall ~X\in\Gamma(A),~m\in
  \Gamma(E_{-1}).
\end{equation}
}

Furthermore, we have
\begin{eqnarray*}
 l_3'(f_0(X+u),f_0(Y+v),f_0(Z+w))&=&c_3'(X,Y,Z)-\Big(\omega(X,Y)(e_1(Z)+c.p.(X,Y,Z)\Big)\\
&& -\big(\omega(X,Y)(w)+\omega(Y,Z)(u)+\omega(Z,X)(v)\big),\\
 f_1(l_3(X+u,Y+v,Z+w))&=&c_3(X,Y,Z)-\big(\omega(X,Y)(w)+\omega(Y,Z)(u)+\omega(Z,X)(v)\big).
\end{eqnarray*}
Thus, we have
\begin{eqnarray*}
&&l_3'(f_0(X+u),f_0(Y+v),f_0(Z+w))-f_1(l_3(X+u,Y+v,Z+w))\\
&=&c_3'(X,Y,Z)-c_3(X,Y,Z)-\Big(\omega(X,Y)(e_1(Z))+c.p.(X,Y,Z)\Big).
\end{eqnarray*}
On the other hand,  by the fact $f_2(u,\cdot)=0$ for all $u\in \Gamma(E_0)$, we have
\begin{eqnarray*}
&&l_2'(f_0(X+u),f_2(Y+v,Z+w))- f_2(l_2(X+u,Y+v),Z+w)+c.p.(X+u,Y+v,Z+w)\\&=&d_\nabla
f_2(X,Y,Z).
\end{eqnarray*}
\emptycomment{
\begin{eqnarray*}
  -\Big(f_2(c_2(X,Y),Z)+c.p.(X,Y,Z)\Big)\\
&&+ \Big(\nabla_X(f_2(Y,w)+f_2(v,Z)+f_2(v,w))-  f_2([X,Y]_A+c_2(X,Y),w)  -
   f_2(\nabla_Xv-\nabla_Yu,Z+w)\\&&+c.p.(X+u,Y+v,Z+w) \Big).
\end{eqnarray*}
}
By \eqref{eq:l3l'3}, we obtain
\begin{eqnarray}\label{eq:iso5}
(c_3-c_3')(X,Y,Z)&=&d_\nabla f_2(X,Y,Z)-\Big(\omega(X,Y)(e_1(Z))+c.p.(X,Y,Z)\Big).
\end{eqnarray}
\emptycomment{
and
\begin{eqnarray}\nabla_Xf_2(Y,w)+\nabla_Yf_2(w,X)=f_2([X,Y]_A+c_2(X,Y),w)-f_2(\nabla_Yw,X)+f_2(\nabla_Xw,Y).
\label{eq:one-more}
 \end{eqnarray}
By \eqref{eq:iso4} and \eqref{eq:iso6}, if $\partial$ is injective,
or surjective, we have
$$
f_2(X,u)=0, \quad\forall~X\in \Gamma(A),~u\in\Gamma(E_0),
$$
which implies that
$$
f_2(c_2(X,Y),Z)+c.p.(X,Y,Z)=0.
$$
Thus, if $\partial$ is injective, or surjective, or $c_2=0$,
}

 Define
$e_2\in\Gamma(\Hom(\wedge^2A,E_{-1}))$ by $e_2(X,Y)=f_2(X,Y)$. By
\eqref{eq:iso3} and \eqref{eq:iso5}, we have
$$
( c_2, c_3)=(c_2',c_3')+D(e_1,e_2).
$$
Therefore, $(c_2,  c_3)$ and $(c_2',c_3')$ are in the same
cohomology class. \qed\vspace{3mm}

Thus, by Lemma \ref{lemma:ext-rep}, Lemma \ref{lemma:ext-cocycle},
Proposition \ref{pro:ext-a-nq} and Proposition \ref{pro:coh},  we have the
following conclusion

\begin{thm}\label{thm:ext-a}
Given a Lie algebroid $A$ and its $2$-term representation up to
homotopy $\huaE$,  the isomorphism classes of the abelian extensions
of $A$ by $\huaE$ are classified by $H^2(A, \huaE)$.
\end{thm}

\begin{rmk}\label{rmk:cohomology}
 It turns out that the natural cohomology group
$H^\bullet(\huaA, \huaE)$ is not enough to include general abelian
extensions even in the case of $L_\infty$-algebras if we  do not add Conditions \eqref{eq:defiext} and \eqref{eq:defiiso}. One needs to go to
a more general version of cohomology group as given in
\cite{lazarev:extension}. In \cite{lazarev:extension}, Lazarev established a one-to-one correspondence between isomorphic classes of extensions of an $L_\infty$-algebra $U$ by an $L_\infty$-algebra $I$ and equivalent classes of $L_\infty$-morphisms from $U$ to the differential graded Lie algebra $ C_{\rm CE}(I,I)[1]$, where $C_{\rm CE}(I,I)$ denotes the Chevalley-Eilenberg complex of the $L_\infty$-algebra $I$.

Recall that in the classical case, extensions of a Lie algebra $\g$ by a $\g$-module $V$ are classified by $H^2(\g,V)$. In this case, there is a
cohomology group in the classical sense. That is, it is the cohomology
of a cochain complex (the Chevalley-Eilenberg complex) $\Sym(\g[1])^* \otimes V$. But already when
dealing with the nonabelian case, there is no representation and no
cohomology in the classical sense. However, we may replace them by studying $L_\infty$-morphisms. Nonabelian extensions of a Lie algebra $\g$ by
a Lie algebra $\frkh$ one-to-one correspond to
  $L_\infty$-morphisms from $\g$ to the Lie $2$-algebra
$\frkh\stackrel{\ad}{\longrightarrow}{\Der(\frkh)}$. 
The connection of the two points of view is as follow:
$k$-cocycles of Chevalley-Eilenberg complex $\Sym((\g[1])^*)\otimes V$ can be
understood as $L_\infty$-morphisms $\g \to \End(V[k-2]) \oplus V[k-1]$
such that their
restriction on $\g\to \End(V[k-2])$ is given by the $\g$ representation
on $V$, and the space of $k$-coboundaries is given by homotopies
between $L_\infty$-morphisms. For extensions of $L_\infty$-algebras, Lazarev replaces 
the right hand side of the $L_\infty$-morphisms by a much
bigger space, namely the space of derivations on the symmetric algebra of the dual of
the $L_\infty$-module. What we see in our version of cohomology group may be viewed as the constant and linear terms.   Therefore, the correct cohomology group corresponding to general
extensions is rather big:  only in the cofibrant case (e.g. when the
kernel of the extension is free), the space is equivalent to (but not necessarily
the same as) the
classical case (see Lemma 2.2 in \cite{lazarev:extension}).

Nevertheless, our version of
cohomology group has its adventage of being very explicit. Moreover, the space of cocycles and
coboundaries has nice geometric meaning as explained in the theory
of representation up to homotopy.  We reserve the generalization of  Lazarev's work to the Lie algebroid context for future
study.
\emptycomment{
Unlike in the classical case, we do not have a full classification
result apparently because $f_2$ controls the deficiency of
bracket-preserving only through $\partial$. It is the case even when
the representation is trivial and when we have Lie $2$-algebras.
Unfortunately, it is not easy to come up with a counter example
either, because $f_2$ needs to satisfy \eqref{eq:one-more}.

This demonstrates the limitation of our naive version of the cohomology
group $H^\bullet (\huaA, \huaE)$ taking simply as the cohomology of
the complex $D: C(\huaA)\otimes \Gamma(\huaE)\to C(\huaA)\otimes
\Gamma(\huaE) $.

In the case of $L_\infty$-algebras, it turns out that
to obtain the full version of one-to-one correspondence, one needs to use a
much bigger cohomology group. Roughly speaking, one needs to replace the space
of endomorphisms of an abelian module (or that of derivations of a general module)
by a much
bigger space involving the symmetric algebra of
the $L_\infty$-module (\cite{lazarev:extension}). These two
cohomologies are equivalent if the kernel of the extension
is free.  At the same time, the concept of an extension also
needs to be taken more general. However, this of course has the cost
that both cocycles and extensions take a much more abstract
appearance.

Thus, we still choose to use this limited version of cohomology, where
cocycles may be written more explicitly. In the end, it turns out that it
is enough for our application in this paper.}

\end{rmk}
We call  the element in $H^2(A, \huaE)$ corresponding to an
extension, the {\bf extension class} of this extension. When the
extension class is $0$, the extension is given by the semidirect
product in the last section. Thus we also have

\begin{thm}\label{thm:ext-b}
Given a Lie algebroid $A$ and its $2$-term representation up to
homotopy $\huaE$, an abelian extension of $A$ by $\huaE$ is
isomorphic to the semidirect product $A\ltimes \huaE$ if and only if
its extension class in $H^2(A, \huaE)$ is trivial.
\end{thm}

Similarly to Theorem \ref{thm:Qreputh}, we have

\begin{thm}\label{thm:Qreputh2cocycle}
Let $\huaE=E_{-1} \oplus E_0$ be a $2$-term graded vector bundle
over $M$ and $A$  a Lie algebroid over $M$. Suppose that there is
a Lie $2$-algebroid structure on $(A\oplus E_0) \oplus E_{-1}$ given
by a degree $1$ homological vector field $Q$,
\begin{eqnarray*}
 &&\CWM\xrightarrow{Q} \Gamma((A\oplus E_0)^*)\xrightarrow{Q} \Gamma( \wedge^2(A\oplus E_0)^*)\oplus
\Gamma(E_{-1}^*)\\
&&\xrightarrow{Q}\Gamma( \wedge^3(A\oplus E_0)^*)\oplus
\Gamma((A\oplus E_0)^*)\otimes\Gamma(E_{-1}^*)\xrightarrow{Q}\cdots.
\end{eqnarray*}
Then this Lie $2$-algebroid is the  abelian extension of $A$ 
if and only if the following conditions are satisfied:
\begin{itemize}
  \item[\rm (1).] the restriction of $Q$ on $C^\infty(M)\oplus
    \Gamma(\wedge^\bullet A^*)$ is exactly given by $d_A$,

\item[\rm (2).] $
Q(\Gamma(E_0^*))\subset\Gamma(E_{-1}^*)\oplus \Gamma(A^*\otimes
E^*_0)\oplus \Gamma(\wedge^2A^*),$

\item[\rm (3).]$
Q(\Gamma(E_{-1}^*))\subset\Gamma(A^*\otimes E_{-1}^*)\oplus
\Gamma(\wedge^2A^*\otimes E^*_0)\oplus\Gamma(\wedge^3A^*). $
\end{itemize}
\end{thm}

Moreover, we have a twisted version of Theorem
\ref{thm:Qsatisfycondition} (which is in the case of $(c_2, c_3)=0$):
\begin{pro}\label{prop:courant-ext}
The underlying NQ manifold of the exact Courant algebroid $T^*[2]T[1]M$ with \v{S}evera class
$[H]\in H^3(M, \R)$ is isomorphic to the one corresponding to the extension of $TM$ by the
coadjoint representation up to homotopy $(T^*M\xrightarrow{id} T^*M,
\nabla, \omega)$ with the extension class
 $$(c_2,c_3)\in \Gamma(\Hom(\wedge^2TM,T^*M) \oplus \Hom(\wedge^3TM,T^*M))$$ given by
\begin{eqnarray}
 \label{eq:c2H}c_2(X,Y)&=&i_{X\wedge Y}H,\\
 \label{eq:c3H}c_3(X,Y,Z)&=&d_\nabla
 c_2(X,Y,Z)=\nabla_Xc_2(Y,Z)-c_2([X,Y],Z)+c.p.(X,Y,Z).
\end{eqnarray}
\end{pro}
\pf The extension Lie 2-algebroid structure is given by
\begin{eqnarray*}
  l_2(X+\xi,Y+\eta)&=&[X,Y]+\nabla_X\eta-\nabla_Y\xi+i_{X\wedge
  Y}H,\\
  l_2(X+\xi,\eta)&=&\nabla_X\eta,\\
  l_3(X+\xi,Y+\eta,Z+\gamma)&=&d_\nabla
  c_2(X,Y,Z)-\Big(\omega(X,Y)(\gamma)+\omega(Y,Z)(\xi)+\omega(Z,X)(\eta)\Big).
\end{eqnarray*}
It fits into the following exact sequence of Lie
 $2$-algebroids,
\begin{equation}\label{eq:ext-courant}
\CD
  0 @>0>>  T^*M @>id_{-1}>> T^*M @>0>> 0 @>0>> 0 \\
  @V 0 VV @V id VV @V 0+id VV @V0 VV @V0VV  \\
  0 @>0>>T^*M  @>\inc>> TM\oplus T^*M  @>\p>> TM@>0>>0.
\endCD
\end{equation}
To see that it corresponds to the exact Courant algebroid with
\v{S}evera class $[H]$, we only need to show that in local
coordinates, their degree 1 homological vector fields are same.
Take the same local coordinates as in Section \ref{sec:courant} and
choose the trivial connection given by these coordinates, that is
$\nabla_{\theta_i} \xi^j=0$ (see Remark \ref{rmk:D}). By  the same
remark, we only need to show that in these coordinates, the dual
$\widetilde{c_2}, \widetilde{c_{3}}$ of $c_2, c_3$ satisfies
$$
\widetilde{c_2}+\widetilde{c_{3}}=\frac{1}{6}\frac{\partial
\phi_{ijk}(q)}{\partial q^l}\xi^i\xi^j\xi^k\frac{\partial}{\partial
p_l}-\frac{1}{6} \phi_{ijk}(q)\xi^i\xi^j\frac{\partial}{\partial
\theta_k}.
$$
 In fact, by \eqref{eq:ctilde}, we have
$
\widetilde{c_2}(\theta_k)(\theta_i,\theta_j)=-\theta_k(c_2(\theta_i,\theta_j))=-\frac{1}{6}
\phi_{ijk}(q)$. Thus, $ \widetilde{c_2}=-\frac{1}{6}
\phi_{ijk}(q)\xi^i\xi^j\frac{\partial}{\partial \theta_k}. $ Now we
have that $c_3=d_\nabla c_2=\frac{1}{6}\frac{\partial
\phi_{ijk}(q)}{\partial q^l}\xi^i\xi^j\xi^kdq^l.$ Thus, by
\eqref{eq:ctilde}, we have $
\widetilde{c_3}(p_l)(\theta_i,\theta_j,\theta_k)=p_l(c_3(\theta_i,\theta_j,\theta_k))=\frac{1}{6}
\frac{\partial \phi_{ijk}(q)}{\partial q^l}, $ which implies that $
\widetilde{c_3}=\frac{1}{6}\frac{\partial \phi_{ijk}(q)}{\partial
q^l}\xi^i\xi^j\xi^k\frac{\partial}{\partial p_l}. $ This completes
the proof. \qed\vspace{2mm}

However we have the following lemma:
\begin{lemma} \label{lemma:trivial-coho}
The cohomology group $H^k(TM, T^*M \xrightarrow{id}T^*M)=0$ for all $k\ge 1$.
\end{lemma}
\pf
Take a $k$-cocycle $(c_k, c_{k+1}) \in \Hom(\wedge^k TM, T^*M) \oplus \Hom (\wedge^{k+1} TM, T^*M)$. Then $D(c_k, c_{k+1})=0$ implies that $d_\nabla c_k = (-1)^k c_{k+1}$.
 Thus $(c_k, c_{k+1})=D(0, (-1)^k c_k)$ is a coboundary.
\qed\vspace{2mm}

Thus, by Theorem \ref{thm:ext-a}, Proposition \ref{prop:courant-ext}, and the above lemma, we conclude that,
\begin{cor}
The exact Courant algebroid $T^*[2]T[1]M$ with \v{S}evera class $[H]$
is isomorphic as NQ manifolds to the standard Courant algebroid with $0$ \v{S}evera class.
\end{cor}
This result is also proven in \cite{LS} using a different method.

We do have an example of non-trivial extension.
\begin{ep} [String Lie 2-algebras] \label{ep:string} {\rm A string Lie $2$-algebra is a $2$-term
  $L_\infty$-algebra $\hat{\frkg}$ with $\hat{\g}_0=\frkg$ a semisimple
  Lie algebra of compact type, $\hat{\g}_{-1}=\R$, and
\begin{eqnarray*}
~ \widehat{\partial}&=&0, \\
\quad l_2((e_1,r_1),(e_2,r_2))&=&([e_1,e_2],0),\\
~l_3((e_1,r_1),(e_2,r_2),(e_3,r_3))&=&(0,\langle [e_1,e_2],e_3 \rangle_{\text{Killing}}),
\end{eqnarray*}
where $e_1,~e_2,~e_3\in\frkg,~r_1,~r_2,~r_3\in\mathbb R$. Then a
string Lie 2-algebra is
a central extension of $\g$ by $\R \to 0$,
\begin{equation}\label{eq:ext-string}
\CD
  0 @>0>>  \R@>id>> \hg_{-1} @>0>> 0 @>0>> 0 \\
  @V 0 VV @V 0 VV @V \widehat{\partial} VV @V0 VV @V0VV  \\
  0 @>0>> 0 @>\inc>> \hg_0=\frkg @>\p>> \frkg@>0>>0,
\endCD
\end{equation}
with the extension class represented by $(0, c_3)$ where
\[ c_3: \wedge^3
\g \to \R \quad \text{ is  given by} \quad  c_3(e_1, e_2, e_3)=\langle [e_1,e_2],e_3
\rangle_{\text{Killing}}. \] It is not hard to see that the cochain
space $C^k(\g, \R
\to 0)$ is precisely the usual cochain space $C^{k+1}(\g, \R)$ for Lie
algebra cohomology.  This relation descends to the cohomology level so
that $H^k(\g, \R\to 0)=H^{k+1}(\g, \R)$, which specially implies that $H^2(\g, \R\to 0)=H^{3}(\g, \R)=\R$. Moreover, the class
represented by $(0, c_3)$ in $H^2(\g, \R\to 0)$ corresponds to the class represented by $c_3$ in
$H^3(\g, \R)$, which is known to be non-zero (actually it is the
generator of $H^3(\g, \R)=\R$).}
\end{ep}

\section{Integration}
We now integrate an abelian extension of a Lie algebroid $A$ by a
2-term representation up to homotopy, $(\huaE, D)$, with the
extension class represented by a 2-cocycle, $(c_2, c_3)$. The
general idea is that we first integrate the representation up to
homotopy $(\huaE, D)$ to a representation up to homotopy $(\huaE,
F_1, F_2)$ of the fundamental Lie groupoid $\huaG$ of $A$. Then we
integrate the extension class $(c_2, c_3)$ into a groupoid extension
class $(C_2, C_3)$. Then we use $F_1, F_2$ and $(C_2, C_3)$ to
construct the extension Lie 2-groupoid and take it as the
integration of the extension Lie 2-algebroid. Notice that both of
the above two integration processes have obstructions (see
\cite[Proposition 5.4]{abad-schaetz:integration-rep-homotopy}, \cite[Theorem
4.7]{abad-schaetz:deform}). Here we use this general idea more as a
guideline to construct the  object integrating  a Courant
algebroid.

In the case of Courant algebroids, the Lie algebroid $A=TM$ is the
tangent bundle.  Thus the fundamental groupoid $\huaG$ is the usual
fundamental groupoid $\Pi_1(M)= \tilde{M} \times \tilde{M}/\pi_1(M)
\rightrightarrows M$,
 where $\tilde{M}$ is the simply connected cover of $M$. When $M$ is
 simply connected, $\Pi_1(M)=M\times M$ is simply the pair groupoid. The representation up to homotopy of $TM$ on $T^*M
 \xrightarrow{id} T^*M$ is  the coadjoint representation up to homotopy.
By \cite[Theorem 3.11]{abad-crainic:rep-homotopy}, any such two
 representations up to homotopy are equivalent. Thus we assume that
 the coadjoint representation  of $TM$ integrates to a
 coadjoint representation up to homotopy $(T^*M\xrightarrow{id} T^*M,
 F_1, F_2)$ of $\Pi_1(M)$.



\subsection{Preliminaries}

Let $\huaG= (G_1\rightrightarrows G_0)$ be a Lie groupoid. We denote
the space of sequences $(g_1,\cdots,g_k)$ of composable arrows (i.e.
$t(g_i)=s(g_{i-1})$) in $ \huaG$ by $\huaG_k$.

\begin{defi}{\rm\cite{abad-crainic:rep-gpd}}
A unital $2$-term representation up to homotopy of a Lie groupoid
 consists of
\begin{itemize}
\item[\rm 1.] A $2$-term complex of vector bundles over $G_0$:
$E_{-1}\stackrel{\partial}{\longrightarrow}E_0$.

\item[\rm 2.] A nonassociative representation $F_1$ on $E_0$ and $E_{-1}$
  satisfying $$ \partial\circ F_1 = F_1\circ \partial, \quad F_1(1_G)=id.$$

\item[\rm 3.] A smooth map $F_2:\huaG_2\longrightarrow\End(V_0,V_{-1})$ such that
\begin{equation}\label{eqn:F fail}
F_1(g_1)\cdot F_1(g_2)-F_1(g_1 g_2)=[\partial,F_2(g_1,g_2)],
\end{equation}
as well as
\begin{equation}\label{eqn:F closed}
F_1(g_1)\circ F_2(g_2,g_3)-F_2(g_1 g_2,g_3)+F_2(g_1, g_2
g_3)-F_2(g_1, g_2)\circ F_1(g_3)=0.
\end{equation}
\end{itemize}
\end{defi}

Given such a representation up to homotopy $(\huaE,F_1,F_2)$, we can
also define a complex to compute the cohomology $H^\bullet(\huaG,
\huaE)$ as in the case of usual representations (see
\cite[Remark 2.8]{abad-schaetz:deform}).  Here we recall the formula
in the case of a 2-term representation: the complex is
\[C(\huaG, \huaE)^n=\oplus_{n=k+l} C^k(\huaG, E_l), \quad \text{where} \; C^k(\huaG, E_l )
 = Maps(\huaG_k, t^*E_l).\]  The differential $D$ is given by
$D= \td + \tF_1 + \tF_2$,
where given $\eta\in C^k(\huaG, E_l)$,
\begin{eqnarray*}
\td(\eta) &=&\partial \circ \eta,\\\tF_1(\eta)(g_1, \cdots,
g_{k+1})&=& (-1)^{k+l}\Big(F_1(g_1)\eta(g_2, \cdots,
g_{k+1})\\
&&+\sum_{i=1}^p(-1)^i\eta(g_1,\cdots,g_ig_{i+1}, \cdots, g_{k+1})
+(-1)^{k+1} \eta(g_1, \cdots, g_k)\Big),
\end{eqnarray*} and
\[\tF_2(\eta)(g_1, \cdots, g_{k+2})=F_2(g_1, g_2) (\eta(g_3, \cdots, g_{k+2})).
\]

In the case of a 2-term representation $E_{-1}\xrightarrow{\partial}
E_0$, a 2-cochain is composed of two terms $(C_2,C_3) \in C^2(G, E_0)
\oplus C^3(G, E_{-1})$.
 The cocycle conditions read
\begin{align} \tF_1(C_2) +\partial\circ C_3 =0, \label{eq:c2-c3-10}\\
\tF_1(C_3)+ \tF_2(C_2)=0. \label{eq:c2-c3-21}
\end{align}
Throughout this paper, unless specifically  mentioned, all
cocycles are normalized, that is, $$\eta(g_1, \dots, g_k)=0,
\quad\mbox{ if one of}~ g_1, \dots, g_k~\mbox{is} ~1_x~\mbox{ for
some}~ x \in G_0.$$

\subsection{Extensions of Lie groupoids} \label{sec:ext-gpd}

First  we recall a classical fact: given a representation $V$ of a group $G$, and a 2-cocycle $C\in C^2(G, V)$, there is a group extension
\[
1\to V \to \widehat{G} \to G\to 1,
\]
where $V$ is viewed as an abelian group with the group operation corresponding to the additive structure of the vector space. When the 2-cocycle is
trivial, $\widehat{G}$ is isomorphic to the semidirect product
$G\ltimes V$.

We would like to establish a similar theory in the Lie 2-groupoid
case and show that the integration of an exact Courant algebroid is such an
extension Lie 2-groupoid (because the Courant algebroid itself can be
realized as an extension Lie 2-algebroid).

   Lie $n$-groupoids can be modelled by certain Kan complexes. However, to describe a Lie 2-groupoid, there is  another method,
 which is much longer to write down, but easier to understand as a
 comparison with Lie groupoids. A Lie 2-groupoid is a groupoid object in
 the 2-category \GpdBibd\; where the space of objects is only a manifold (but not a general Lie groupoid). Here \GpdBibd\; is the 2-category with Lie groupoids as objects,
  Hilsum-Skandalis (HS) bimodules as morphisms, and isomorphisms of HS bimodules as 2-morphisms. Thus the category of manifolds embeds into this 2-category by viewing
  a manifold $M$ as a trivial groupoid $M\rightrightarrows M$ which only has identity arrows. The equivalence of this description with that in terms of Kan complexes is given in \cite{z:tgpd-2}.
  A special sort of Lie 2-groupoid is a groupoid object in the
  2-category of \Gpd \; with  the space of objects a manifold,  where
  \Gpd\; is a sub-2-category of \GpdBibd \; containing only strict groupoid morphisms as morphisms.
  We call such Lie 2-groupoid semistrict Lie 2-groupoid. Since the Lie 2-groupoid integrating Courant algebroid that we construct is an example of semistrict Lie 2-groupoids,
  we describe this concept explicitly below. Here we borrow largely
  from \cite[Section 7]{baez:2gp}.

\begin{defi} \label{def:semi-2gpd}
A semistrict Lie $2$-groupoid consists of:
\begin{itemize}
  \item[$\bullet$] a smooth manifold $G_0$, which is the set of
  objects $x,y,z,\cdots,$

 \item[$\bullet$] a smooth manifold $G_1$, which is the set of
  $1$-morphisms $g,h,\cdots$. For a $1$-morphism $g:x\longrightarrow
  y$, we write $\alpha(g)=x,~\beta(g)=y$. For another $1$-morphism $h:y\longrightarrow
  z$, we write their composition as $hg:x\longrightarrow z$.

   \item[$\bullet$] a Lie groupoid $\mathscr{G}: G_2\rightrightarrows G_1$,
     where $G_2$ is the set of $2$-morphisms $\phi$, $\phi'\cdots$. For any $2$-morphism $\phi:g\Longrightarrow h$,
   where $g,h:x\longrightarrow y$ are $1$-morphisms, the source and target
   maps $s,t$ are given by $s(\phi)=g,~t(\phi)=h$. The composition
   in this groupoid is usually called vertical multiplication, and
   denoted by $\cdot_\ve.$ We require $\beta s = \beta t$ and $\alpha s
   = \alpha t$.

\item[$\bullet$] For all objects $x,y,z\in G_0$, there is a Lie groupoid morphism
$\mathscr{G}\times_{\alpha s, G_0, \beta s} \mathscr{G} \to
\mathscr{G}$, which is called horizontal multiplication and denoted
by $\cdot_\h$, i.e. for $\phi:g\Longrightarrow h:x\longrightarrow y$
and $\phi^\prime:g^\prime\Longrightarrow h^\prime:y\longrightarrow
z$, we have
$$
\phi^\prime\cdot_\h\phi: g^\prime g\Longrightarrow h^\prime
h:x\longrightarrow z,
$$
or, in terms of a diagram,
$$\xymatrix@C+2em{
 z &
  \ar@/_1pc/[l]_{g'}_{}="0"
  \ar@/^1pc/[l]^{h'}^{}="1"
  \ar@{=>}"0";"1"^{\phi'}
 y }\cdot_\h\xymatrix@C+2em{
 y &
  \ar@/_1pc/[l]_{g}_{}="0"
  \ar@/^1pc/[l]^{h}^{}="1"
  \ar@{=>}"0";"1"^{\phi}
  x }=\xymatrix@C+2em{
 z &
  \ar@/_1pc/[l]_{g'g}_{}="0"
  \ar@/^1pc/[l]^{h'h}^{}="1"
  \ar@{=>}"0";"1"^{\phi'\cdot_\h\phi}
  x. }$$
 \item[$\bullet$] for any $x\in G_0$, there is an identity $1$-morphism and an identity $2$-morphism, which we both denote by $1_x$.

\item[$\bullet$]  a Lie groupoid contravariant morphism $\inv: \mathscr{G} \to \mathscr{G}$,
\end{itemize}
and the following natural isomorphisms
\begin{itemize}
  \item[$\bullet$] the associator $a_{(g_1,g_2,g_3)}:(g_1\cdot_\h g_2)\cdot_\h g_3\longrightarrow g_1\cdot_\h (g_2\cdot_\h
  g_3)$. 

 \item[$\bullet$] the left and right unit $l_g:1_{\beta(g)}\cdot_\h g\longrightarrow
 g$ and $r_g:g\cdot_\h 1_{\alpha(g)}\longrightarrow g$,

\item[$\bullet$] the unit and counit $i_g:1_{\beta(g)}\longrightarrow g\cdot_\h
\inv(g)$ and $e_g:\inv(g)\cdot_\h  g\longrightarrow 1_{\alpha(g)}$
\end{itemize}
which are such that the following diagrams commute:
\begin{itemize}
  \item[$\bullet$] the {\bf pentagon identity } for the associator
{\footnotesize\[
 \xymatrix{ & (g_1\cdot_\h g_2)\cdot_\h(g_3\cdot_\h g_4)\ar[dr]^{a_{g_1,g_2, g_3\cdot_\h g_4}}&  \\
((g_1\cdot_\h g_2)\cdot_\h g_3)\cdot_\h g_4\ar[ur]^{a_{(g_1\cdot_\h
g_2),g_3,g_4}}\ar[dr]_{a_{g_1,g_2,g_3}\cdot_\h 1_{g_4}}&&Q\\
&(g_1\cdot_\h (g_2\cdot_\h g_3))\cdot_\h
g_4\stackrel{a_{g_1,g_2\cdot_\h g_3,g_4}}{\longrightarrow}
g_1\cdot_\h ((g_2\cdot_\h g_3)\cdot_\h g_4)\ar[ur]^{1_{g_1}\cdot_\h
a_{g_2,g_3,g_4}}&}
\]}
where $Q=g_1\cdot_\h
(g_2\cdot_\h (g_3\cdot_\h g_4))$.
\item[$\bullet$] the {\bf triangle identity} for the left and right
unit:
\[
 \xymatrix{
( g_1\cdot_\h 1_{\alpha(g_1)})\cdot_\h
g_2\ar[rr]^{a_{g_1,1_{\alpha(g_1)},g_2}}\ar[dr]^{r_{g_1}\cdot_\h
1_{g_2}}&&g_1\cdot_\h(1_{\alpha(g_1)}\cdot_\h g_2)
\ar[dl]^{1_{g_1}\cdot_\h l_{g_2}}\\
&g_1\cdot_\h g_2& }
\]
\item[$\bullet$]the {\bf first zig-zag identity}:

{\footnotesize{
\[
 \xymatrix{
&(g\cdot_\h \inv(g))\cdot_\h g\stackrel{a_{g,\inv(g),g}}{\longrightarrow}g\cdot_\h(\inv(g)\cdot_\h g)\ar[dr]^{1_g\cdot_\h e_g}&\\
1_{\beta(g)}\cdot_\h g\ar[dr]^{l_g}\ar[ur]^{i_g\cdot_\h 1_g}&&g\cdot_\h 1_{\alpha(g)}\\
&g\ar[ur]^{r_g^{-1}}&}
\]
}}

\item[$\bullet$] the {\bf second zig-zag identity}:

\footnotesize{
\[
 \xymatrix{
&\inv(g)\cdot_\h (g\cdot_\h \inv(g))\stackrel{a_{\inv(g),g,\inv(g)}}{\longrightarrow}(\inv(g)\cdot_\h g)\cdot_\h\inv(g)\ar[dr]^{e_g\cdot_\h 1_{\inv(g)}}&\\
\inv(g)\cdot_\h 1_{\beta(g)}\ar[dr]^{r_{\inv(g)}}\ar[ur]^{1_{\inv(g)}\cdot_\h i_g}&&1_{\alpha(g)}\cdot_\h \inv(g).\\
&\inv(g)\ar[ur]^{l_{\inv(g)}^{-1}}&}
\]
}
\end{itemize}
\end{defi}
For simplicity, we denote a semistrict Lie 2-groupoid by
$G_2\rightrightarrows G_1\rightrightarrows G_0$. In the special case
where $a_{g_1,g_2,g_3}$,~$l_g$,~$r_g,$~$i_g,$~$e_g$ are all identity
isomorphisms, we call such a Lie 2-groupoid a {\bf strict Lie
 $ 2$-groupoid.} If $G_0$ is a point, we obtain the concept of a {\bf semistrict Lie
  $2$-group} \cite{baez:2gp,shengzhu1}.

Any vector bundle  $E$ can be viewed as an abelian Lie
groupoid with the source and the target both the projection to the
base and multiplication the pointwise addition. Similarly, we have

\begin{ex}\label{ex:abelian2g}{\rm
Any $2$-term complex of vector bundles
$\huaE:E_{-1}\stackrel{\partial}{\longrightarrow} E_0$ has an
``abelian'' strict Lie $2$-groupoid structure, which we denote by
$E_\bullet$. First,
  we have an action groupoid $E_0\rtimes_{G_0} E_{-1} \rightrightarrows E_0$
  where $E_{-1}$ acts on $E_0$ by $$u \cdot m= u + \partial m.$$ Furthermore, the pointwise addition of $E_i$ gives horizontal multiplication of the Lie $2$-groupoid,
  that is
\begin{equation}\label{eq:ab-2-gpd}
(u, m) \cdot_\h (v, n) = (u+ v, m+n).
\end{equation}
Moreover $inv(u, m)=(-u, -m)$, $1_p=(0_p, 0_p)$ for a point $p$ on the base.}
\end{ex}

An {\bf abelian extension} of a Lie groupoid
$\huaG=G_1\rightrightarrows G_0$ by a 2-term
representation $\huaE$ is a short exact sequence of Lie 2-groupoids
with the left term $E_\bullet$ viewed as an abelian Lie 2-groupoid.
Now we explain what an exact sequence of Lie 2-groupoid is
(but only in our very special case)\footnote{There should be a more
general notation of exact sequence of Lie 2-groupoids using
generalized morphisms allowing higher morphisms. It should include a
Kan fibration as an example. On the other hand, our definition here
is a special case of Kan fibration.}.  In our special case, all the
Lie 2-groupoid morphisms we mention here are strict, namely they
respect all the structure maps strictly without further 2-morphisms.
Given  a Lie 2-groupoid $\hat{G}_\bullet$ and a Lie groupoid $\huaG$
with $\hat{G}_0=G_0$, a 2-groupoid morphism $\phi_\bullet:
\hat{G}_\bullet \to \huaG$ with $\phi_0=id_{G_0}$ is surjective if
$\phi_1$ is a surjective submersion (then this implies that $\phi_2$
is a surjective submersion).  Given another Lie 2-groupoid
$\hat{H}_\bullet$ with $\hat{H}_0=\hat{G}_0$, a 2-groupoid morphism
$\iota_\bullet: \hat{H}_\bullet \to \hat{G}_\bullet$ with
$\iota_0=id_{G_0}$ is injective if $\iota_1$ and $\iota_2$ are
embeddings.
 The image $\im(\iota_\bullet)$ is naturally defined as the image 2-groupoid under $\iota_\bullet$. The kernel $\ker(\phi_\bullet)$ is made up by $G_0$,
 and the subsets of $\hat{G}_{1}$ and $\hat{G}_2$ which maps to
 $\{1_x, x\in G_0\}$ under $\phi_\bullet$. Since the identity of
 $G_\bullet$ is strict and $\phi_{1}$, $\phi_2$ are
 surjective submersions, $\ker(\phi_\bullet)$ is a Lie 2-groupoid. We call the short sequence
 $\hat{H}_\bullet \xrightarrow{\iota_\bullet} \hat{G}_\bullet \xrightarrow{\phi_\bullet}\huaG$ exact if $\iota_\bullet$ is injective,
  $\phi_\bullet$ is surjective, and $\ker(\phi_\bullet)=\im(\iota_\bullet)$ as Lie 2-groupoids.
\[
\xymatrix{ &H_2 \ar@{=>}[d] \ar[r]^{\iota_2} & \hat{G}_2 \ar@{=>}[d]  \ar[r]^{\phi_2} & G_1 \ar@{=>}[d] \\
1\ar[r] & H_1 \ar@{=>}[d] \ar[r]^{\iota_1} & \hat{G}_1 \ar@{=>}[d] \ar[r]^{\phi_1} & G_1 \ar@{=>}[d] \ar[r] & 1.\\
&H_0=G_0 \ar[r]^{\iota_0} & \hat{G}_0=G_0 \ar[r]^{\phi_0} & G_0 }
\] We have the following proposition
  which can be viewed as the global version of Lemma \ref{lemma:ext-cocycle}:
(however, we shall not expect a classification result as in
Theorem \ref{thm:ext-a} with our current version of groupoid cohomology because even in the case of groups this version needs to be refined for the classification result
to hold. See \cite[Section 2]{wz:int} and \cite{SP}).

\begin{pro}\label{prop:gpd-ext}
Given a $2$-term representation up to homotopy $(\huaE, F_1, F_2)$
of a Lie groupoid $\huaG = G_1 \rightrightarrows  G_0$ and a
$2$-cocycle $(C_2, C_3) \in C^2(\huaG, \huaE)$, there is a Lie
$2$-groupoid structure on $G_1 \times_{G_0} E_0 \times_{G_0} E_{-1}
\rightrightarrows G_1 \times_{G_0} E_0 \rightrightarrows G_0$ which
is an extension of $\huaG$ and fits into the exact sequence
\[
\xymatrix{ &E_0\times_{G_0} E_{-1} \ar@{=>}[d] \ar[r]^{\iota_2\qquad} & G_1 \times_{G_0} E_0 \times_{G_0} E_{-1} \ar@{=>}[d]  \ar[r]^{\qquad\qquad\phi_2} & G_1 \ar@{=>}[d] \\
1\ar[r] & E_0 \ar@{=>}[d] \ar[r]^{\iota_1} & G_1 \times_{G_0} E_0 \ar@{=>}[d] \ar[r]^{\phi_1} & G_1 \ar@{=>}[d] \ar[r] & 1\\
&G_0 \ar[r]^{\iota_0=id_{G_0}} & G_0 \ar[r]^{\phi_0=id_{G_0}} & G_0
}
\] with natural inclusion $\iota_\bullet$ and natural projection $\phi_\bullet$. The Lie $2$-groupoid structure of the left term is abelian as
in Example \ref{ex:abelian2g}. The Lie $2$-groupoid structure of the
middle term is semistrict and given by the following data: The
source map $s$ and the target map $t$ are given by \be\label{s t}
s(g,\xi,m)=(g,\xi),\quad t(g,\xi,m)=(g,\xi+\partial m), \ee and
$\alpha,\beta$ (see the second item of Definition \ref{def:semi-2gpd}) are given by
$$
\alpha(g,\xi)=\alpha(g),\quad\beta(g,\xi)=\beta(g),
$$
for any $(g,\xi)\in G_1\times_{G_0} E_0$.

 The vertical
multiplication $\cdot_\ve$ is given by
$$
(h,\eta,n)\cdot_\ve(g,\xi,m) =(g,\xi,m+n),\quad \mbox{where}~
h=g,\eta=\xi+\partial m.
$$
The horizontal multiplication $\cdot_\h$ of objects is given by
\be\label{m o} (g_1,\xi)\cdot_\h (g_2,\eta)=(g_1
g_2,\xi+F_1(g_1)(\eta) + C_2(g_1, g_2)), \ee the horizontal
multiplication $\cdot_\h$ of morphisms is given by \be\label{m m}
(g_1,\xi,m)\cdot_\h (g_2,\eta,n)=(g_1 g_2,\xi+F_1(g_1)(\eta)
+C_2(g_1, g_2),m+F_1(g_1)(n)). \ee The  associator
$$
a_{(g_1,\xi),(g_2,\eta),(g_3,\gamma)}:\big((g_1,\xi)\cdot_\h(g_2,\eta)\big)\cdot_\h(g_3,\gamma)\longrightarrow
(g_1,\xi)\cdot_\h\big((g_2,\eta)\cdot_\h(g_3,\gamma)\big)
$$
is given by {\footnotesize\begin{eqnarray}
\nonumber&&a_{(g_1,\xi),(g_2,\eta),(g_3,\gamma)} =\\
 &&\label{associator}(g_1 g_2
g_3,\xi+F_1(g_1)(\eta)+F_1(g_1 g_2)(\gamma) + C_2(g_1, g_2)+
C_2(g_1g_2, g_3),F_2(g_1,g_2)(\gamma)-C_3(g_1, g_2, g_3)).
\end{eqnarray}}
The inverse map $\inv$ is given by \be
\inv(g,\xi)=(g^{-1},-F_1(g^{-1})(\xi)-C_2(g^{-1}, g)), \ee and \be
\inv(g, \xi, m)= (g^{-1},-F_1(g^{-1})(\xi)-C_2(g^{-1}, g),
-F_1(g^{-1})(m)). \ee The
identity $1$-morphisms are $(1_x,0)$ and the identity $2$-morphisms are $(1_x,0,0)$.\\
 The unit $i_{(g,\xi)}:(1_{\beta(g)},0)\longrightarrow
(g,\xi)\cdot_\h \inv(g,\xi)$ is given by \be\label{unit}
i_{(g,\xi)}=(1_{\beta(g)},0,-F_2(g,g^{-1})(\xi)+C_3(g,g^{-1},g)).
\ee All the other natural isomorphisms are identity
isomorphisms. Moreover, two $2$-cocycles representing the same class in
$H^2(\huaG, \huaE)$ give rise to isomorphic Lie $2$-groupoid structures.
If the extension class $(C_2, C_3)=0$, we call such an extension a
{\bf semidirect product} of $\huaG$ with its representation up to
homotopy $(\huaE, F_1, F_2)$.
\end{pro}

\pf First we verify  the Lie 2-groupoid structure of the middle
term. The verification is similar to the proof of \cite[Theorem
3.7]{shengzhu1}.

Firstly, it is
straightforward to verify that the multiplication $\cdot_\h$ is a groupoid
morphism. Then we compute that
\begin{eqnarray*}
\big((g_1,\xi)\cdot_\h(g_2,\eta)\big)\cdot_\h(g_3,\gamma)=(g_1g_2g_3,\xi+F_1(g_1)(\eta)+C_2(g_1,g_2)+F_1(g_1g_2)(\gamma)+C_2(g_1g_2,g_3)),\end{eqnarray*}
and \begin{eqnarray*}
 (g_1,\xi)\cdot_\h\big((g_2,\eta)\cdot_\h(g_3,\gamma)\big)=\big(g_1g_2g_3,\xi+F_1(g_1)(\eta+F_1(g_2)(\gamma)+C_2(g_2,g_3))+C_2(g_1,g_2g_3)\big).
\end{eqnarray*}
By \eqref{eqn:F fail}, $a$ defined by \eqref{associator} is the
associator iff
$$
F_1(g_1)C_2(g_2,g_3)-C_2(g_1g_2,g_3)+C_2(g_1,g_2g_3)-C_2(g_1,g_2)+\partial
C_3(g_1,g_2,g_3)=0.
$$ This is exactly \eqref{eq:c2-c3-10}--one of the conditions of the
closedness of  $(C_2,C_3)$.

The naturality of  the associator $a$  is the following commutative
 diagram:
{\footnotesize\[ \xymatrix{
\big((g_1,\xi)\cdot_\h(g_2,\eta)\big)\cdot_\h(g_3,\gamma)\ar[d]\ar[r]^{a}&(g_1,\xi)\cdot_\h\big((g_2,\eta)\cdot_\h(g_3,\gamma)\big)\ar[d]\\
\big((g_1,\xi+\partial m)\cdot_\h(g_2,\eta+\partial
n)\big)\cdot_\h(g_3,\gamma+\partial k)\ar[r]^{a}&(g_1,\xi+\partial
m)\cdot_\h\big((g_2,\eta+\partial n)\cdot_\h(g_3,\gamma+\partial
k)\big). }
\] }To see that $a$ is a natural isomorphism,  we need to show that
 \be\label{left}
a_{(g_1,\xi+\partial m),(g_2,\eta+\partial n),(g_3,\gamma+\partial
k)}\cdot_\h \Big(\big((g_1,\xi,m)\cdot_\h (g_2,\eta,n)\big)\cdot_\h
(g_3,\gamma,k)\Big) \ee is equal to \be\label{right}
\Big((g_1,\xi,m)\cdot_\h \big((g_2,\eta,n)\cdot_\h
(g_3,\gamma,k)\big)\Big)\cdot_\h
a_{(g_1,\xi),(g_2,\eta),(g_3,\gamma)}.\ee By straightforward
computations, we  obtain that (\ref{left}) is equal to
{\footnotesize$$ \big(g_1 g_2 g_3,\xi+F_1(g_1)(\eta)+F_1(g_1
g_2)(\gamma),m+F_1(g_1)(n)+F_1(g_1
g_2)(k)+F_2(g_1,g_2)(\gamma+\partial k)-C_3(g_1, g_2, g_3)\big),
$$}
and (\ref{right}) is equal to {\footnotesize$$ \big(g_1 g_2
g_3,\xi+F_1(g_1)(\eta)+F_1(g_1 g_2)(\gamma),m+F_1(g_1)(n)+F_1(g_1)
F_1(g_2)(k)+F_2(g_1,g_2)(\gamma)-C_3(g_1, g_2, g_3)\big).
$$}
 Hence (\ref{left}) is equal to (\ref{right}) by (\ref{eqn:F fail}).

By (\ref{eqn:F fail}), \eqref{eq:c2-c3-10}, $F_1(1_x)=id$   and the
fact that our cocycles are normalized, we have
\begin{eqnarray*}
  (g,\xi)\cdot_\h\inv(g,\xi)=\big(1_{\beta (g)},-\partial F_2(g,g^{-1})(\xi)+\partial
C_3(g,g^{-1},g)\big).
\end{eqnarray*}
 Thus the unit given by (\ref{unit}) is
indeed a morphism from $(1_{\beta(g)},0)$ to $(g,\xi)\cdot_\h
\inv(g,\xi)$.

To show the naturality of the unit, we need to prove
$$
\big((g,\xi,m)\cdot_\h\inv(g,\xi,m)\big)\cdot_\ve
i_{(g,\xi)}=i_{(g,\xi+\partial m)},
$$
i.e. the following commutative diagram:

$$
\xymatrix{
  &  (1_{\beta(g)},0)\ar[dr]^{i_{(g,\xi)}} \ar_{i_{(g,\xi+\partial m)}}[dl]\\
(g,\xi+\partial m)\cdot_\h \inv(g,\xi+\partial m)
&&\ar[ll]_{\quad\qquad\qquad~(g,\xi,m)\cdot_\h\inv(g,\xi,m)}(g,\xi)\cdot_\h\inv(g,\xi)
}
$$
 This follows from
 $$
F_2(g,g^{-1})(\partial m)=F_1(g)\cdot F_1(g^{-1})(m)-F_1(g\cdot
g^{-1})(m)=F_1(g)\cdot F_1(g^{-1})(m)-m,
 $$ which is a special case of (\ref{eqn:F fail}).

Since $F(1_x)=id$ and our cocycle $C_2+C_3$ is normalized,  we have
$$(1_{\beta(g)},0)\cdot_\h (g,\xi)=(g,\xi),\quad (g,\xi)\cdot_\h(1_{\alpha(g)},0)=
(g,\xi).$$ Hence the left unit and the right unit can also be taken
as the identity isomorphisms.

 The counit
$e_{(g,\xi)}:\inv(g,\xi)\cdot_\h (g,\xi)\longrightarrow
(1_{\alpha(g)},0)$ can be taken as the identity morphism since we
have
$$
\inv(g,\xi)\cdot_\h (g,\xi)=\big(g^{-1},-F_1(g^{-1})(\xi) -
C_2(g^{-1}, g)\big)\cdot_\h (g,\xi)=(1_{\alpha(g)},0).
$$
Now we need to show various coherence conditions between 2-morphisms.
Here we only give the proof of the  pentagon identity. The others can be
proved in similar fashions and we leave them to the readers. The
pentagon identity is equivalent to
\begin{eqnarray*}
&&a_{(g_1,\xi),(g_2,\eta),(g_3,\gamma)\cdot_\h(g_4,\theta)}\cdot_\ve
a_{(g_1,\xi)\cdot_\h(g_2,\eta),(g_3,\gamma),(g_4,\theta)}=\\
&&\big((g_1,\xi)\cdot_\h
a_{(g_2,\eta),(g_3,\gamma),(g_4,\theta)}\big)\cdot_\ve
a_{(g_1,\xi),(g_2,\eta)\cdot_\h(g_3,\gamma),(g_4,\theta)}\cdot_\ve
\big(a_{(g_1,\xi),(g_2,\eta),(g_3,\gamma)}\cdot_\h(g_4,\theta)\big)
\end{eqnarray*}
The condition \eqref{eq:c2-c3-10} implies that these elements can be
vertically multiplied.  Then by straightforward computations, the left
hand side is equal to {\footnotesize\begin{align*} \Big(g_1g_2
g_3g_4,\xi+F_1(g_1)(\eta)+F_1(g_1 g_2)(\gamma)+F_1(g_1
g_2g_3)(\theta) +C_2(g_1,g_2)+ C_2(g_1g_2, g_3)+C_2(g_1g_2g_3, g_4), \\
F_2(g_1 g_2, g_3)(\theta)+F_2(g_1,
g_2)\big(\gamma+F_1(g_3)(\theta)+C_2(g_3,g_4)\big) -C_3(g_1g_2, g_3,
g_4) - C_3(g_1, g_2, g_3g_4) \Big),
\end{align*}}
and the right hand side is equal to {\footnotesize\begin{eqnarray*}
&&\Big(g_1 g_2 g_3 g_4,\xi+F_1(g_1)(\eta)+F_1(g_1
g_2)(\gamma)+F_1(g_1g_2 g_3)(\theta) +C_2(g_1,g_2) + C_2(g_1g_2,
g_3)+C_2(g_1g_2g_3, g_4),\\&&F_2(g_1, g_2)(\gamma)+F_2(g_1, g_2
g_3)(\theta)+F_1(g_1)\circ F_2(g_2, g_3)(\theta)-F_1(g_1)C_3(g_2,
g_3, g_4)\\
&& - C_3(g_1, g_2g_3, g_4) - C_3(g_1, g_2, g_3) \Big).
\end{eqnarray*}}
By \eqref{eqn:F closed} and \eqref{eq:c2-c3-21}, they are equal.

It is not hard to verify that the natural inclusion $\iota_\bullet$ is injective and the natural projection of $\phi_\bullet$ is surjective.
 Moreover, $\ker(\phi_\bullet)=\im(\iota_\bullet)=E_\bullet$. Thus we indeed obtain an
 extension.

Suppose that $(C_2, C_3)=(C'_2, C'_3)+ D(B_1, B_2)$. Then the groupoid
automorphism $\Phi=(\Phi_0, \Phi_1)$ of $G\times E_0\times E_1 \Rightarrow
G\times E_0$,
\[\Phi_0(g, \xi)= (g, \xi+B_1(g)), \quad \Phi_1(g, \xi, m)=(g, \xi+B_1(g), m),
\]
gives an isomorphism of Lie 2-groupoids. The only technical point to
notice is that the groupoid morphisms $\cdot_\h \circ (\Phi\times \Phi)$ and
$\Phi \circ (-\cdot_\h -)$ are not exactly identical but differ by a
2-morphism $\Phi_2: G\times E_0 \times G \times E_0 \to G \times E_0
\times E_1$ given by $\Phi_2((g, \xi), (h, \eta))=(gh, \xi+\eta, B_2(g,
h))$. \qed

\begin{rmk}\label{rk:diff}
It is proven in \cite{LS} that the differentiation of
$\hat{G}$ is exactly the Lie $2$-algebroid corresponding to an Courant
algebroid.  Here we remark how to do the differentiation from the point of view of extensions.
The systematic way to differentiate  a Lie $n$-groupoid to an NQ
manifold is described in \cite{severa:diff} in the language of
graded manifolds. We describe briefly the differentiation
inspired by this work (using explicit words in usual differential
geometry), and postpone the detailed calculation to future studies.

Recall the differentiation of a Lie groupoid $t,s:
H_1\rightrightarrows H_0$ to a Lie algebroid. The Lie algebroid is
$\ker Ts|_{H_0}$ containing tangent vectors of $H_1$ on $H_0$
vanishing along the source map.   Similarly, we obtain a graded
vector bundle
\[\ker T\alpha|_{G_0} \oplus \ker Ts|_{G_0}[1] = (A\oplus E_0) \oplus E_{-1}[1],\]
where $A$ is the Lie algebroid of $G$. Now we explain how to obtain
the Lie $2$-algebroid structure \eqref{eq:ext-a-structure} on this
graded vector bundle. The anchor $\rho$ of the Lie $2$-algebroid is
induced by the anchor of $A$.

We notice that the formula for the horizontal multiplication
\eqref{m o} is exactly the same as the formula for a usual extension
of groupoid by a representation and a $2$-cocycle. This implies the
second formula of \eqref{eq:ext-a-structure}. Moreover, we notice
that there is an (nonassociative) action of $G_1\oplus E_0$ on
$E_{-1}$ given by the horizontal multiplication:
\[ (g, \xi) \cdot m :=pr_{E_{-1}}\big( (g, \xi, 0) \cdot_h (1_{s(g)}, 0, m) \big)= F_1(g)(m).
\]This implies the third formula of \eqref{eq:ext-a-structure}.
The term $l_3$ is more difficult to explain, but at least in our case of
Courant algebroids, it is determined by $l_2$ since $\partial: E_{-1} \to E_0 $ is injective.
\end{rmk}

\subsection{Application to integration of Courant algebroids}
Now we are ready to integrate the Courant algebroid $TM\oplus T^*M$
with \v{S}evera class $[H]$. The integrating Lie 2-groupoid is
simply the extension Lie 2-groupoid of $\Pi_1(M)$ by its coadjoint
representation $T^*M \xrightarrow{id}T^*M$ and a certain 2-cocycle
$(C_2, C_3)$ in $C^2(TM,T^*M \xrightarrow{id}T^*M)$ (which turns out
to be not important). More explicitly, the Lie 2-groupoid is modeled
on the action Lie groupoid $(\Pi_1(M) \times_{M} T^*M )\rtimes_{M}
T^*M \rightrightarrows \Pi_1(M)\times_{M} T^*M$ with $T^*M $ acts on
$\Pi_1(M) \times_{M} T^*M$ by addition on $T^*M$. The structure maps
are given as in Proposition \ref{prop:gpd-ext}.

We now give a description of this Lie 2-groupoid by Kan complex using the correspondence in
\cite[Section 2.3]{z:tgpd-2} using the notation in Proposition \ref{prop:gpd-ext}. The 0-th
 level $X_0=M$ is simply the base of the
Courant algebroid. The first level is
\[X_1=\Pi_1(M)\times_{t, M} T^*M (=\hat{G}_1), \]
with $d_1=\beta$ and $d_0=\alpha$ and $s_0$ the natural embedding
$M\to \Pi_1(M)\times_{t, M} T^*M$. The second level is
\[X_2=\big( \Pi_1(M) \times_{t, M, s} \Pi_1(M) \big)\times_{t\circ pr_l \times t
  \circ pr_r \times t \circ pr_l, M^{\times 3}} T^*M^{\times 3},\]
such that for a typical element $(\gamma_{0,1}, \gamma_{1,2}, \xi_{x_0},
\xi_{x_1}, m_{x_0})\in X_2$,
\begin{align*}
d_0(\gamma_{0,1}, \gamma_{1,2}, \xi_{x_0},
\xi_{x_1}, m_{x_0}) &=(\gamma_{1,2}, \xi_{x_1}),  \\
d_1(\gamma_{0,1}, \gamma_{1,2}, \xi_{x_0},
\xi_{x_1}, m_{x_0}) &=(\gamma_{0,1} \gamma_{1,2}, \xi_{x_0} +
C_2(\gamma_{0,1}, \gamma_{1,2}) + F_1(\gamma_{0,1}) (\xi_{x_1})+ id( m_{x_0})),  \\
d_2(\gamma_{0,1}, \gamma_{1,2}, \xi_{x_0}, \xi_{x_1}, m_{x_0})
&=(\gamma_{0,1}, \xi_{x_0}).
\end{align*}
Then in general $X$ is determined by the first three levels,
$X = Cosk_3 Sk_3 (\Lambda[3, 0](X), X_2, X_1, X_0)$ (see \cite[Section 2.3]{z:tgpd-2}).
That is $X_n$ is a fibre product made up by $X_2$'s, $X_1$'s and $X_0
$'s. In this special case, it is every easy to see that $X$ is a Lie
2-groupoid since the differential from $T^*M$ to $T^*M$ is an
isomorphism so that
$X_2$ is totally determined by its images under $d_0$, $d_1$ and
$d_2$. More explicitly there is a simplicial manifold $Y_\bullet$ with
\[
\begin{split}
 Y_n= &\Pi_1(M)^{\times n} \times_{M^{\times \big(^{n+1}_2\big)}} T^*M^{\times \big(^{n+1}_2\big)} \\
=&\{ (\gamma_{0,1}, \gamma_{1,2}, \dots, \gamma_{n-1, n}; \dots ,\xi^{i,j}, \dots): 0\le i < j \le n, \gamma_{i,i+1} \in \Pi_1(M) \; \text{is represented by } \\
& \text{a path from $x_i$ to $x_{i+1}$, and $\xi^{i, j} \in
T^*_{x_i}M$ arranged in dictionary order.} \}
\end{split}
\] One should imagine each element as the dimensional-1-skeleton of a $n$-polygon in $M$ with each edge attached with a cotangent vector at the end.
The face and degeneracy maps are naturally given by
\[
d_k(\gamma_{0,1}, \gamma_{1,2}, \dots, \gamma_{n-1, n}; \dots, \xi^{i,j}, \dots) =
(\dots, \gamma_{k-1, k} \cdot \gamma_{k, k+1}, \dots; \dots, \hat{\xi}^{i,k}, \dots, \hat{\xi}^{k, j}, \dots ),
\]
\[s_k(\gamma_{0,1}, \gamma_{1,2}, \dots, \gamma_{n-1, n}; \dots, \xi^{i,j}, \dots) =
(\dots, \gamma_{k-1, k}, 1_{x_k}, \gamma_{k, k+1}, \dots; \dots, \tilde{\xi}^{i,j}, \dots
),
\] with $\tilde{\xi}^{i,j}=\xi^{i,j}$ for $i<j\le k$, $\tilde{\xi}^{k,k+1}=0$, $\tilde{\xi}^{i, j} = \xi^{i-1, j-1}$ for $k<i<j$,
$\tilde{\xi}^{i,j}=\xi^{i, j-1}$ for $i\le k \le j-1$. Since $Y_\bullet$ is
determined by its 1-skeleton, it is clearly a Lie 2-groupoid.
Moreover, regardless of the cocycle $(C_2, C_3)$, we have $X_\bullet
\cong Y_\bullet$ as a simplicial manifolds since both  are
determined by their 1-skeleton. More precisely, the isomorphism
$X_2\cong Y_2$ given by
\[ \big(\gamma_{01}, \gamma_{12}, \xi_{x_0}, \xi_{x_1}, m_{x_0} \big) \mapsto
\big(\gamma_{0,1}, \gamma_{1,2}; \xi_{x_0}, \xi_{x_0}+C_2(\gamma_{0,1},
\gamma_{1,2})+F_1(\gamma_{0,1})(\xi_{x_1})+m_{x_0} ,\xi_{x_1} \big), \]
together with the natural identification of $X_i=Y_i$ when $i=0,1$,
induce an isomorphism of the simplicial manifolds  $X_\bullet
\cong Y_\bullet$.
 If we take a local neighborhood of $Y_0=M$ in
$Y_n$, we arrive at the local Lie $2$-groupoid
$\mathcal{T} \mathcal{M}$ in
\cite[Theorem 1]{LS}, which  differentiates to the standard Courant algebroid
$(T^*[2]T[1]M, [H])$ by this theorem. Thus we have,
\begin{thm}
An exact Courant algebroid $(T^*[2]T[1]M, [H])$ as a Lie $2$-algebroid
integrates to the semidirect product Lie $2$-groupoid of $\Pi_1(M)$
with its coadjoint representation up to homotopy
$T^*M\xrightarrow{id}T^*M$, regardless of the \v{S}evera class
$[H]$.
\end{thm}

\begin{rmk}[Comparison with other works]
In  \cite{LS}, the authors also construct a global version using
 the pair groupoid rather than the fundamental groupoid as we do in our approach. But
there is no fundamental difference. This global Lie $2$-groupoid
built upon the pair groupoid is also the integration object of Mehta
and Tang \cite[{\rm(1.3)}]{MT}.  We choose the fundamental groupoid in
hope that we would achieve a more universal object, possibly with a
universal symplectic structure. However, it seems that we need to go
further (to ``fundamental $2$-groupoid'') to achieve this universal
object, since our current object is not source $2$-connected. This direction is
further studied in a
recent article \cite{MTb}. They apply the general Sullivan-\v{S}evera-Getzler-Henriques integration
procedure (studied in details for the $L_\infty$-algebroid case in
\cite{SS}) and build a universal infinite dimensional $2$-groupoid, called
the {\rm LWX} $2$-groupoid (named after Liu-Weinstein-Xu). They apply the procedure to $T^*M$ but not the Lie
$2$-algebroid corresponding to $T^*[2]T[1]M$ directly, most likely
basing on a certain hiding equivalence between them.  They claim that the relation
between the {\rm LWX} $2$-groupoid and the finite dimensional one is
not yet entirely
clear, though we think a suitable projection should be expected.
\end{rmk}

\def\cprime{$'$} \def\cprime{$'$} \def\cprime{$'$} \def\cprime{$'$}

\end{document}